\documentclass[11pt]{article}
\usepackage[a4paper, margin=2.5cm]{geometry}
\usepackage{multirow}
\newcommand{\hs}{\hspace*{0.5cm}}
\usepackage{hyperref}
\usepackage{graphicx,amsmath,amssymb}
\usepackage{enumerate}
\usepackage{psfrag}
\usepackage[usenames,dvipsnames]{xcolor}
\usepackage{caption}
\usepackage{color}   
\usepackage{changepage}
\usepackage{soul}
\usepackage{cite}
\makeatletter \renewcommand\@biblabel[1]{#1.~} \makeatother
\makeatletter \def\@cite#1#2{${}^{\, \mbox{\scriptsize #1\if@tempswa , #2\fi}}$} \makeatother
\let\oldbibliography\thebibliography
\renewcommand{\thebibliography}[1]{%
  \oldbibliography{#1}%
  \setlength{\itemsep}{-1pt}%
}

\newcommand{\ag}{{\rm ag}}

\newcommand{\s}{{\bf s}}
\newcommand{\anc}{{\sl 1}} 
\newcommand{\vac}{{\rm vac}} 
\newcommand{\bst}{{\rm bst}} 
\newcommand{\EQ}{\begin{equation}}
\newcommand{\EE}{\end{equation}}
\newcommand{\EQA}{\begin{eqnarray}}
\newcommand{\EEA}{\end{eqnarray}}

\usepackage{lineno}

\begin{document}
\captionsetup[figure]{labelfont={bf},name={Fig.},labelsep=space}
\noindent 
{\Large \bf Vaccination shapes evolutionary trajectories of SARS-CoV-2 }
\vspace{0.5cm} 
\\
Matthijs Meijers${}^{\, \rm a}$, Denis Ruchnewitz${}^{\, \rm a}$, Marta \L uksza${}^{\rm b}$, Michael L\"assig${}^{\, \rm a, \ast}$ 
\vspace{0.5cm} 
\\ 
{\small
${}^{\, \rm a}$ Institute for Biological Physics, University of Cologne, Z\"ulpicherstr.~77, 50937 K\"oln, Germany \\
${}^{\, \rm b}$ Tisch Cancer Institute, Departments of Oncological Sciences and Genetics and Genomic Sciences, Icahn School of Medicine at Mount Sinai, New York, NY, USA \\
${}^\ast$ To whom correspondence should be addressed. Email: mlaessig@uni-koeln.de 
}
\vspace{0.5cm} 

\noindent
\subsection*{Abstract}
The large-scale evolution of the SARS-CoV-2 virus has been marked by rapid turnover of genetic clades. New variants show intrinsic changes, notably increased trans\-missibility, as well as antigenic changes that reduce the cross-immunity induced by previous infections or vaccinations\cite{Ozono2021,Meng2021,Mlcochova2021,Harvey2021}. How this functional variation shapes the global evolutionary dynamics has remained unclear. Here we show that selection induced by vaccination impacts on the recent antigenic evolution of SARS-CoV-2; other relevant forces include intrinsic selection and antigenic selection induced by previous infections. We obtain these results from a fitness model with intrinsic and antigenic fitness components. To infer model parameters, we combine time-resolved sequence data\cite{Shu2017}, epidemiological records\cite{owidcoronavirus,coronaUSA}, and cross-neutralisation assays\cite{Planas2021c,Planas2021b,GarciaBeltran2022}. This model accurately captures the large-scale evolutionary dynamics of SARS-CoV-2 in multiple geographical regions. In particular, it quantifies how recent vaccinations and infections affect the speed of frequency shifts between viral variants. Our results show that timely neutralisation data can be harvested to identify hotspots of antigenic selection and to predict the impact of vaccination on viral evolution. 

\subsection*{Introduction}

Two classes of molecular adaptation have been observed in the evolution of SARS-CoV-2 \linebreak to date. Multiple mutations carry intrinsic changes of viral functions, such as increasing the binding affinity to human receptors\cite{Ozono2021}, the efficiency of cell entry\cite{Meng2021,Mlcochova2021}, or the stability of viral proteins\cite{Wrobel2020,Zeng2021}. Other mutations, referred to as antigenic changes, decrease the neutralizing activity of human antibodies\cite{Harvey2021,Planas2021b,Planas2021c,GarciaBeltran2022}, thereby reducing the immune protection against secondary infections\cite{Khoury2021,Feng2021}. The strains that inherit a given mutation define a clade of the evolving viral population. Several of these molecular changes had drastic evolutionary and epidemiological impact, inducing global turnover of viral clades and concurrent waves of the pandemic. Over the last two years, three genetic variants and their associated clades successively gained global prevalence: Alpha~($\alpha$) from March to June in 2021, Delta~($\delta$) from June to December in 2021 and Omicron~($o$) in 2022. These were named Variants of Concern (VOCs) by the World Health Organization\cite{WHO_VOC}; other VOCs gained temporary regional prevalence. Several studies reported fitness advantages of VOCs inferred from epidemiological trajectories and comparative functional studies\cite{Davies2021,Dhar2021,Kepler2021,Mlcochova2021,Ulrich2022}. Importantly, however, the evolutionary impact of antigenic changes is time-dependent, because it depends on previously acquired population immunity: a larger amount of previous infections or vaccinations increases the global fitness advantage of an antigenic escape mutation. 
Specifically, multi-strain epidemiological models and simulations suggest that vaccinations can favour the emergence of escape variants\cite{Grenfell2004,Rella2021,SaadRoy2021,Lobinska2022} and influence the turnover of circulating clades\cite{Luksza2014,Wen2022}; effects of this kind have been reported for some clades of human influenza\cite{wen2018}. In the case of SARS-CoV-2, pandemic infection and massive vaccination programs, with a global count of 4.5 billion vaccinations and $>$200 million confirmed cases in 2021\cite{owidcoronavirus}, have built up partial population immunity, but its feedback on viral evolution has not been quantified.
This leads to the central question of this paper: 
what is the impact of vaccination and infection rates on the turnover of SARS-CoV-2 clades? To address this question, we infer a data-driven fitness model for SARS-CoV-2 variants with distinct components of intrinsic fitness  and antigenic fitness by vaccination and infection. 

\subsection*{Results}
\paragraph*{Trajectories and speed of clade turnover} 

As a first step, we map the evolutionary trajectories of the three global clade shifts in the last two years. To track circulating clades, we analyse a set of $>$5M quality-controlled SARS-CoV-2 sequences obtained from the GISAID database\cite{Shu2017}. We assign these sequences to genetic clades using a standard set of amino acid changes\cite{Outbreak}; then we infer time-dependent clade frequencies from strain counts smoothened over a period of $\sim$30 days (Methods). To obtain accurate, time-resolved data, we record frequency trajectories at the level of regions (countries and US states). 
 Including all regions satisfying uniform criteria of 
data availability (Methods), we obtain frequency trajectories of the $\anc -\alpha$, the $\alpha-\delta$, and the $\delta-o$ shift for 11, 16, and 14 regions, respectively. Here, 1 denotes the set of clades circulating prior to $\alpha$, including the wild type (wt) and the early 614G mutation in the spike protein. Fig.~1a shows trajectories of the ancestral and the invading clade for the $\alpha-\delta$ and $\delta-o$ shifts in Italy; trajectories for all regions of this study are reported in Fig.~S1 and~S2. 

Assuming that large-scale frequency shifts of viral clades are adaptive processes, we can infer the underlying selective force from frequency trajectories. Specifically, the fitness difference (selection coefficient) between invading and ancestral clades takes the form 
\EQ
\hat{s}(t) = \frac{\rm d}{{\rm d}t} \log \frac{x_{\rm inv}(t)}{x_{\rm anc} (t)},
\label{eq:s_hat}
\EE
where $x_{\rm inv} (t)$ and $x_{\rm anc} (t)$ are the corresponding frequencies (here and below, empirical selection coefficients inferred from frequency trajectories are marked by a hat). We note that this relation is independent of other co-circulating clades (Methods). In Fig.~1b, we show time-resolved, regional selection coefficients of the invading and ancestral clade for the $\alpha-\delta$ and $\delta-o$ shifts. These data reveal two opposing trends: During the $\alpha-\delta$ shift, selection increases with time in 16 of 16 regions. Conversely, selection driving the $\delta-o$ shift decreases with time in 12 of 14 regions. Compared to a reference of time-independent selection,  the $\alpha-\delta$ shift runs at an accelerating speed, the $\delta-o$ shift at a decelerating speed. The time dependence of selection is statistically significant ($P < 10^{-15}$ for $\alpha -\delta$, $P < 10^{-5}$ for $\delta-o$; two-sided Wald test). In contrast, the earlier $\anc - \alpha$ shift does not show a significant signal of time-dependent selection ($P>0.01$, Fig.~S3). In what follows, we will relate this pattern to feedback of vaccination on viral evolution.

\begin{figure*}[h!]
\centering
\includegraphics[width=\linewidth]{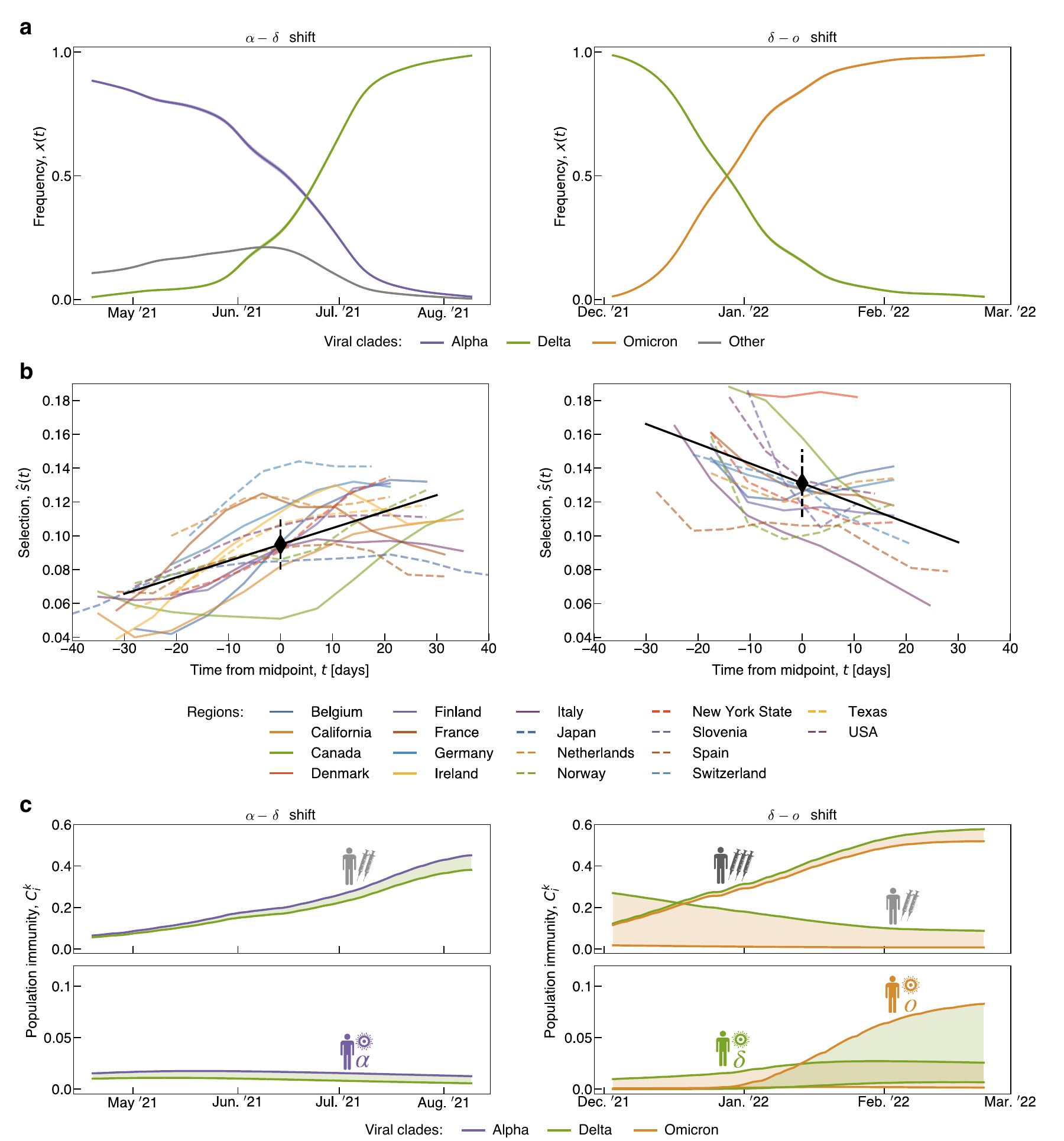}
\caption*{\footnotesize {\bf Fig. 1}:
{\bf Evolutionary, epidemiological, and immune tracking of SARS-CoV-2.} Time-dependent trajectories are shown for the clade shift from $\alpha$ to $\delta$ (left column) and from $\delta$ to $o$ (right column). 
 {\bf (a)}~Observed frequency trajectories of relevant clades, $x_i (t)$, for the clade shifts in Italy.
 {\bf (b)}~Empirical selection coefficient (fitness difference) between invading and ancestral clade, $\hat s (t)$, for all regions. Selection trajectories are derived from the frequency trajectories of (a) and plotted against time counted from the midpoint. Summary statistics: cross-region linear regression (black solid line), cross-region average (black diamond), and rms~cross-region variation of selection (black dashed line). 
 {\bf (c)} Population immunity functions of the ancestral and invading variant, $C_{\rm anc}^k (t)$ and $C_{\rm inv}^k (t)$, in relevant immune channels $k$ for Italy (coloured lines). Cross-immunity differences in a given channel, $C_{\rm inv}^k (t) - C_{\rm anc}^k (t)$,  are highlighted by shading (colours indicate which variant receives a fitness advantage). See Figs.~S1--S3 for tracking of all shifts in all regions and reporting of rms statistical errors. 
 }
\label{fig:antigenic}
\end{figure*}

\paragraph*{Cross-immunity trajectories}

Cross-immunity induced by a primary infection against subsequent infections by related patho\-gens is routinely tested by neutralisation assays, which measure the minimum antiserum concentration required to neutralise the second antigen. Relative, inverse concentrations are reported as serum dilution titers; here we use logarithmic titer values, $T$ (with base~2). 
For SARS-CoV-2, recent work\cite{Planas2021c,Planas2021b,GarciaBeltran2022,Straten2022,Wilks2022} has established a matrix of titers, $T_i^k$, measuring neutralisation of variant $i$ in immune channel $k$ (Fig.~2a, Table S1). Here and below, immune channels label primary challenges inducing specific antisera, including infections by different variants ($k = \alpha, \delta, o, \dots$), as well as primary and booster vaccinations ($k = \vac, \bst$; titers shown here are for mRNA vaccines). Together, these data provide a first, coarse-grained cross-immunity landscape of SARS-CoV-2. Infection-induced cross-immunity titers are maximal when primary infection and secondary challenge are by the same variant
(Fig.~2a). Similarly, titers induced by primary vaccination, $T_i^\vac$, are maximal against strains from the ancestral clade, which contains the strain used for vaccination\cite{Polack2020}. 
Differences of neutralisation titers, $\Delta T_{ij}^k = T_i^k - T_j^k$, measure differences in functional antibody binding between strains of different variants; evolved titer reductions are also referred to as antigenic advance. Notably, each of the global clade shifts observed to date, $\anc-\alpha$, $\alpha-\delta$, and $\delta-o$, has decreased neutralisation by vaccination, i.e., generated antigenic advance, $\Delta T_{\anc \alpha}^\vac, \Delta T_{\alpha \delta}^\vac, \Delta T_{\delta o}^\vac > 0$ (Fig.~2b). 
Moreover, the in-vivo concentration of neutralising antibodies decays exponentially with time after immunisation\cite{Iyer2020,Israel2022}. This translates into a linear titer reduction, $T_i^k (\Delta t) = T_i^k - \Delta t/\tau$, with an estimated half life $\tau \sim 65$ days (Methods). 

\begin{figure*}[t!]
\centering
\includegraphics[width=\linewidth]{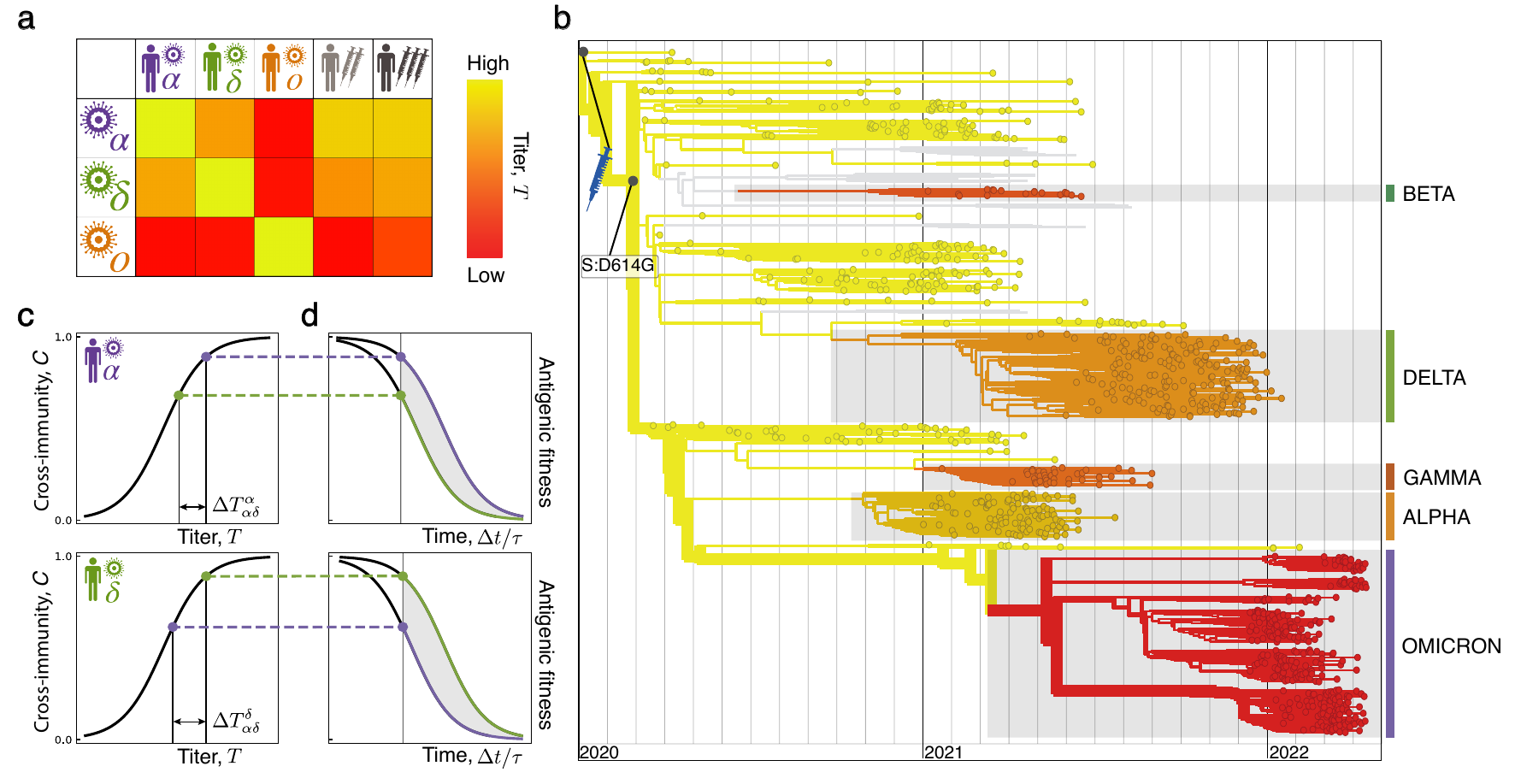}
\caption*{\footnotesize  {\bf Fig.~2: Cross-neutralisation and antigenic fitness.}
{\bf (a)} Neutralisation titers, $T_i^k$, of human antisera induced by different primary challenges (columns: infection by strains from clade $\alpha$, $\delta$, $o$, mRNA primary and booster vaccination) assayed against different test strains (rows: strains from clade $\alpha$, $\delta$, $o$); see Table~S1. 
{\bf (b)} Strain tree of SARS-CoV-2 with lineages $i$ colored by vaccine neutralisation titers, $T_i^\vac$. WHO Variants of Concern are marked by bars. The ancestral clade ($\anc$) has the highest neutralisation (yellow); the successive clade shifts $\anc-\alpha$, $\alpha-\delta$, and $\delta-o$ decrease neutralisation, i.e., induce antigenic advance (see text). 
{\bf (c)} The cross-immunity, $c_i^k$, induced by a primary immunisation in channel $k$ (top: infection by $\alpha$, bottom: infection by $\delta$) against a secondary infection of clade $i$ (blue dot: $\alpha$, red dot: $\delta$) is a Hill function of the neutralising titer, $T_i^k$ (ref.~[\citen{Khoury2021,Feng2021}], Methods). Cross-immunity decreases with increasing antigenic advance $\Delta T_{ik}^k$ (bars).
{\bf (d)} Cross-immunity decays with time after primary immunisation, $\Delta t$ (in units of the characteristic decay time $\tau$; ref.~[\citen{Iyer2020,Israel2022}]). According to the fitness model, cross-immunity induces a proportional antigenic fitness cost; the resulting time-dependent selection coefficient (fitness difference) between clades is marked by shading. 
}
\label{fig:schematic}
\end{figure*}

Importantly, recent work for SARS-CoV-2 has also shown that neutralisation titers predict the cross-immunity $c_i^k$, defined as the relative drop of secondary infections in human cohorts. Specifically,  $c_i^k = H (T_i^k)$ is a Hill function\cite{Khoury2021,Feng2021} (Fig.~2c, details are given in Methods), consistent with the underlying biophysics of antibody-antigen binding and with results for other viral pathogens\cite{Coudeville2010,Dunning2016,Rotem2018,Meijers2021}. The post-immunisation decay of antibody concentration induces a decay of cross-immunity, $c_i^k (\Delta t) = H (T_i^k - \Delta t /\tau)$. Together, cross-immunity depends in a predictable, nonlinear way on neutralisation titer and on time since primary immunisation. Fig.~2 shows two examples of this pattern. Primary infection by an $\alpha$ strain induces a high cross-immunity against other $\alpha$ strains and a reduced cross-immunity against $\delta$ strains ($c_\alpha^\alpha > c_\delta^\alpha$) (Fig.~2c, top). Both factors decrease by antibody decay; their difference has a maximum at an intermediate time since primary infection (Fig.~2d, top). Infection by a $\delta$ strain induces cross-immunity factors of opposite ranking ($c_\delta^\alpha < c_\delta^\delta$) and similar decay (Fig.~2cd, bottom). 

To track population immunity over time, we combine these cross-immunity factors with infection and vaccination data. In each region, we record cumulative fractions of immunised individuals, $y_k (t)$, in each channel $k$ (clade-specific infections, primary and booster vaccinations; see Figs. S1 and~S2). Their derivatives $\dot y_k (t)$ are the rates of new immunisations in channel $k$. Clade-specific infection data are obtained by multiplying the total rate of new infections reported in each region with the simultaneous viral clade frequencies $x_k (t)$ (Fig.~1a). In the regions included in our analysis, vaccination has been predominantly by mRNA vaccines (Methods). By weighting with the time-dependent cross-immunity factors $c_i^k (\Delta t)$, we infer the population cross-immunity against clade $i$ by immunisation in channel~$k$, 
\EQ
{C}_i^k (t) = \int^{t} c_i^k (t-t') \, \dot y_k (t') \, dt'. 
\label{C}
\EE
In Fig.~1c, we plot the cross-immunity trajectories relevant for the $\alpha-\delta$ and $\delta-o$ shifts in Italy; trajectories for all regions are reported in Figs. S1 and~S2. The $\alpha-\delta$ shift shows sizeable and increasing immunity induced by primary vaccination, while infection-induced immunity remains small. During the $\delta-o$ shift, immunity by primary vaccination declines, while booster- and infection-induced immunity components increase. During the earlier $\anc-\alpha$ shift, population cross-immunity is still small in most regions. We conclude that the joint dynamics of new immunisations and antibody decay can produce complex and opposing cross-immunity patterns.

\paragraph*{Inference of intrinsic and antigenic selection} 

To quantify the feedback of cross-immunity on viral evolution, we use a minimal, computable fitness model, 
\EQ
f_i (t) = f_i^0 - \sum_k \gamma_k C_i^k (t), 
\label{f} 
\EE
where $f_i (t)$ is the absolute fitness, or epidemic growth rate, of a viral strain. Fitness is proportional to the log of the effective reproductive number, $f_i (t) = \tau_0^{-1}\log R_i (t)$, where $\tau_0$ denotes the infectious period (Methods). Here, we write fitness as the sum of a time-independent intrinsic component, $f_i^0$, and of time-dependent antigenic components, $f_i^k (t) = - \gamma_k C_i^k (t)$ (Methods). Each component is proportional to the corresponding cross-immunity factor $C_i^k (t)$ with a weight factor $\gamma_k$ for each immune channel $k$. Hence, selection is generated by  cross-immunity differences  between competing strains (shading in Fig.~1c and Fig.~2c). This type of fitness model has been established for predictive evolutionary analysis of human influenza\cite{Luksza2014,Morris2018,Huddleston2020} and is grounded in multi-strain epidemiological models\cite{Gog2002}. The minimal fitness model does not account for differences in cross-immunity between human hosts (for example, through differences in immunodominance\cite{Lipsitch2019}) and for correlations between multiple prior infections (antigenic sin\cite{Cobey2017}). 

For SARS-CoV-2, we compute fitness at the level of variants, neglecting fitness differences between strains within a clade. Similarly, we evaluate cross-immunity at the level of variant-specific prior infection and of primary and booster vaccination, using the trajectories $C_i^k (t)$ calculated above (Fig.~1c). 
To compare model and data, we compute the fitness difference between invading and ancestral strain for each regional trajectory: $s(t) = f_{\rm inv} (t) - f_{\rm anc} (t) = s_0 + s_\ag (t)$, where intrinsic selection, $s_0$, and antigenic selection, $s_\ag (t) = \sum_k s_k (t)$, are given by equation~(\ref{f}). Then we decompose the model-based selection trajectories into mean and change, $s (t) =  \langle s \rangle + \Delta s (t)$ (brackets denote time averages over the trajectory for a given region). The empirical trajectories $\hat s (t)$ are decomposed in the same way (Fig.~1b). Cross-region selection differences, measured by the rms~deviation of $\langle \hat s \rangle$, reflect inhomogeneous conditions of contact limitations, surveillance, geography, and population structure that are not included in the minimal model. In a given region, however, variants compete under more homogeneous conditions. Therefore, 
we infer antigenic selection from the regional selection change, $\Delta \hat s (t)$. We use a minimal model with just 3 antigenic parameters: a uniform $\gamma_\vac$ for vaccination and boosting (downweighted by a factor $a$ in the $\delta - o$ shift to account for double infections\cite{Bates2022}) and a uniform $\gamma_k = b \gamma_\vac$ for all infection channels $k$ (upweighted by a factor $b$ to correct for relative underreporting; Methods). We infer maximum-likelihood (ML) values of these parameters by calibrating computed and empirical trajectories, $\Delta s (t)$ and $\Delta \hat s (t)$, for the $\alpha-\delta$ and $\delta-o$ shifts. 
 The intrinsic selection coefficients, $s_0$, are then obtained as the time-independent part of selection. 
Details of the inference procedure are given in Methods; ML model parameters and selection coefficients for all clade shifts are reported in Table~S2 and~S3. 

\newpage
\vspace{-3cm}
\begin{figure*}[t]
\centering
\includegraphics[width=\linewidth]{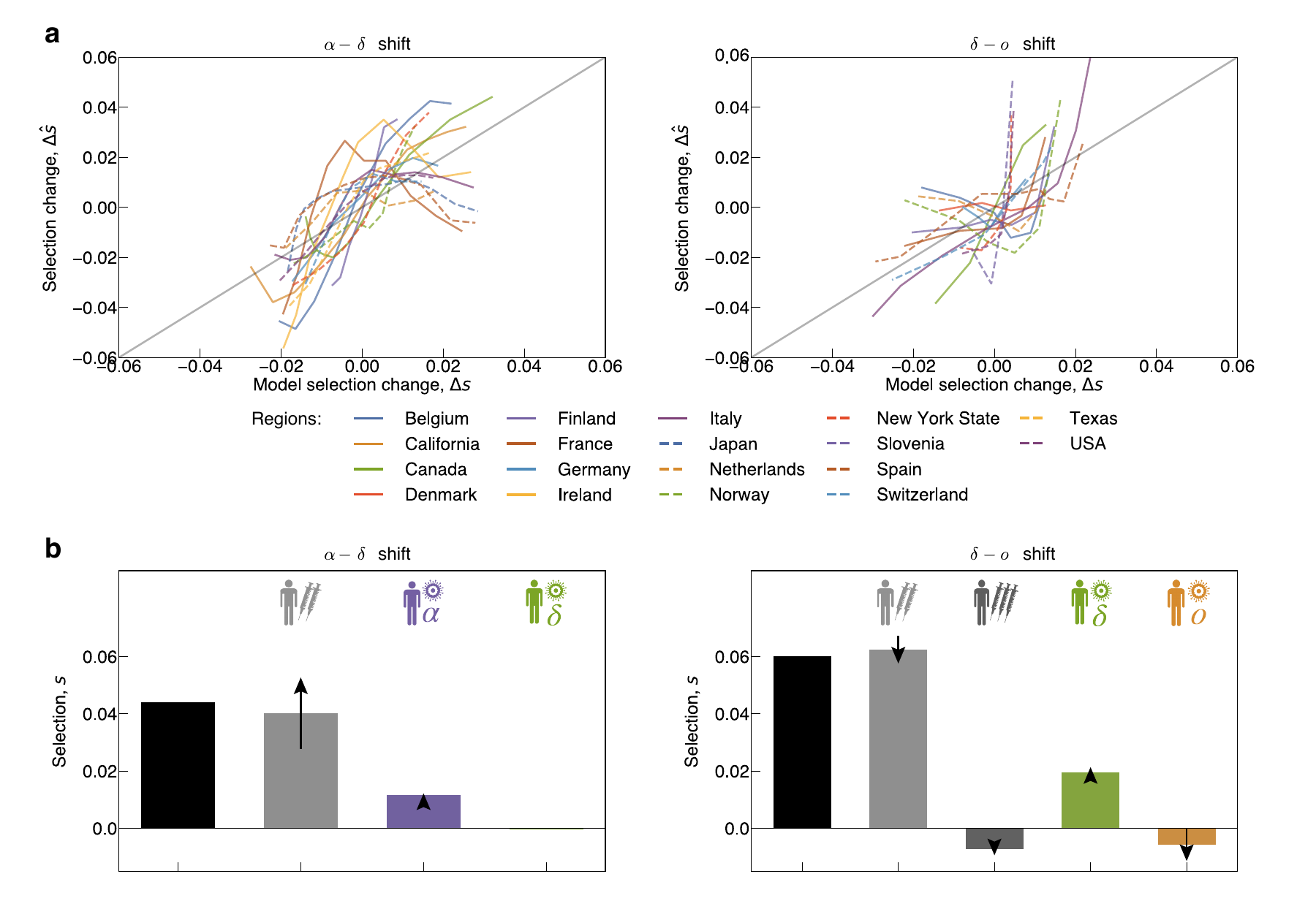}
\caption*{\footnotesize {\bf Fig. 3}:
{\bf Antigenic and intrinsic selection drive SARS-CoV-2 evolution.} We compare empirical selection trajectories and model predictions for the clade shift from $\alpha$ to $\delta$ (left column) and from $\delta$ to $o$ (right column). 
 {\bf (a)}~Empirical selection change, $\Delta \hat s$, obtained from the selection trajectories of Fig.~1a are plotted against model predictions, $\Delta s$, for all regions. Rms statistical errors are reported in Figs.~S1--S2.
 {\bf (b)}~Breakdown of fitness model components. Intrinsic selection coefficients (black) and antigenic selection coefficients in marked immune channels (coloured), as inferred from the ML fitness model (bars: region- and time-averaged value for each crossover; arrows: region-averaged rms~temporal change, $\langle (\Delta s)^2 \rangle^{1/2}$, with marked direction; confidence intervals are given in Table~S3). 
 }
\label{fig:antigenic}
\end{figure*}

In Fig.~3a, we plot $\Delta s$ from the ML fitness model against the corresponding empirical selection change, $\Delta \hat s$, obtained from the trajectories of Fig.~1b. We obtain a remarkable data compression: for most regions, the antigenic fitness computed from equation (\ref{f}) reproduces the empirical fitness changes (intrinsic selection drops out of this comparison). This can be further quantified: the covariance between data and 
ML model, $\langle \Delta s \Delta \hat s \rangle$, 
explains $\sim 50$\% of the empirical variance of selection, $\langle (\Delta \hat s)^2 \rangle$; this level of covariance is found on average and in most individual regions. A detailed comparison of data and model trajectories, $\Delta \hat s (t)$ and $\Delta s (t)$, for all regions is shown in Figs. S1 and~S2. 
 As a control, the model predicts only small selection change for the $\anc - \alpha$ shift, consistent with the weak time dependence of the empirical selection trajectories (Fig.~S3).
We conclude that time-dependent cross-immunity explains the time-dependence of selection governing $\mbox{SARS-CoV-2}$ variant shifts.

\paragraph*{Impact of vaccination and infection on evolution}

From the ML fitness model, we obtain a breakdown of intrinsic and antigenic selection components relevant for each clade shift. Intrinsic selection is strong and positive in all three major clade shifts, with average selection coefficients $s_0 = 0.05 - 0.08$,
consistent with strong functional differences observed between the $\alpha$, $\delta$, and $o$ variants\cite{Mlcochova2021,Yuan2022} (Fig.~3b, Table S3). Antigenic selection becomes equally strong in the $\alpha-\delta$ and $\delta-o$ shifts. Its two main components, vaccination- and infection-induced selection, are statistically significant parts of the fitness model, partial models with only one component have a strongly reduced posterior likelihood (differences in model complexity are accounted for by a Bayesian information criterion; see Methods and Table~S2). 

Vaccination induces cross-immunity differences between variants, resulting in positive antigenic selection of average strength $s_{\rm vac} = 0.04$ in the $\alpha-\delta$ shift and $s_{\rm vac} = 0.06$ in the $\delta-o$ shift (Fig.~3b, Table S3). These selection coefficients quantify the evolutionary impact of primary SARS-CoV-2 vaccination: they measure the relative increase in effective reproduction number of the invading variant by partial escape from vaccination-induced immunity ($\tau_0 s_\vac = R_{\rm inv} / R_{\rm anc} -1$).
Vaccination-induced antigenicity also explains the observed time-dependence of selection (Fig.~1b, Figs.~S1 and~S2): $s_\vac$ increases during the $\alpha-\delta$ shift because of increasing vaccination levels, but decreases during the $\delta-o$ shift because vaccination-induced immunity fades. 
In both shifts, primary vaccination generates the dominant components of antigenic selection (Fig.~3b).
 Booster vaccinations have increased breadth; they induce higher neutralisation $T_\delta^\bst, T_o^\bst$ and reduced antigenic advance  $\Delta T_{\alpha \delta}^\bst, \Delta T_{\delta o}^\bst$ compared to primary vaccinations\cite{Gruell2022,GarciaBeltran2022,Planas2021c,Hachmann2022} (Fig.~2a, Table~S1). Hence, booster vaccinations generate higher cross-immunity but weaker selection for antigenic escape (Fig.~1c, Fig.~4ab). 
 The net effect of boosters in the $\delta-o$ shift is opposite to that of primary vaccinations: we infer a negative selection coefficient $s_\bst = -0.01$. This is because boosters remove cross-immunity differences and antigenic selection generated by the preceding primary vaccination (Fig.~3b).

Infection-induced antigenic selection increased in net strength from 0.01 in the $\alpha-\delta$ shift to 0.03 in the $\delta-o$ shift. Notably, it always contains components of opposite sign: primary infections by the ancestral clade generate positive selection, while infections by the invading clade generate negative selection. This frequency-dependent negative feedback acts to prolong the coexistence of ancestral and invading clade. Together, antigenic selection can produce complex but computable patterns of time dependence. 

 These results require careful interpretation. They show that vaccination and previous infections induced sizeable antigenic selection on circulating SARS-CoV-2 variants and modulated the speed of successive clade shifts. However, antigenic selection did not cause or prevent any of these shifts, because intrinsic functional changes generated sizeable fitness advantages of the invading variants independently of population immunity. The breakdown of selection given in Fig.~3b applies to the set of regions accessible to our analysis; the relative weights of vaccination- and infection-induced selection components are expected to be different in other regions. The availability of comparable data precludes a fully global model-based analysis. An additional, model-free inference of selection in regions with low vaccination coverage is given in Methods. Most importantly, the fitness model and our data analysis do not predict any simple relation between vaccination coverage and speed of evolution. This is because cross-immunity channels are  correlated: fewer vaccinations lead to more infections, generating buildup of cross-immunity in other channels and complex long-term effects. 

\begin{figure*}[h!]
\centering
\includegraphics[width=\linewidth]{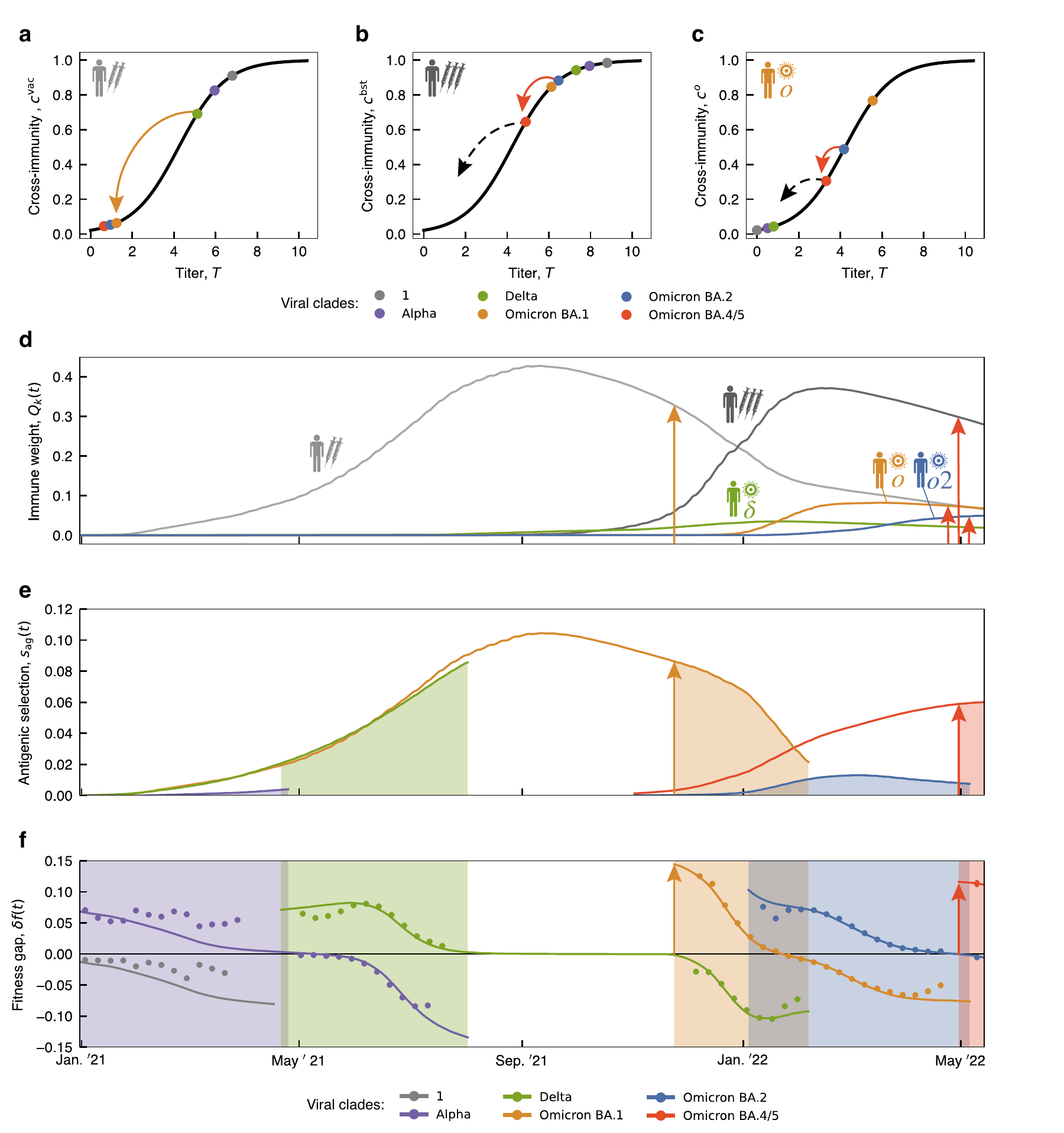}
\caption*{\footnotesize {\bf Fig.~4: Antigenic landscapes and immune weights generate selection hotspots.}
(a) Antigenic landscapes. Cross-immunity factors of major and recent clades (colored dots) are plotted against neutralisation titers in different immune channels: 
{\bf (a)}~primary vaccination, 
{\bf (b)}~booster vaccination, 
{\bf (c)} $o$-induced infection.
 {\bf (d)}~Long-term immune weight trajectories of different channels, $Q_k (t)$. 
 {\bf (e, f)}~Long-term fitness trajectories of major and recent clades. Periods of clade shifts are highlighted by shading.
 {\bf (e)}~Antigenic selection between successive variants, $s_\ag (t)$. 
 Model-based trajectories for each variant pair are shown up to the end of the corresponding clade shift. 
 {\bf (f)}~Time-dependent fitness gap, $\delta f (t)$ (Methods). Model-based trajectories for each variant~$i$ (lines) are shown from in the time interval where $0.01 \leq x_i (t) \leq 0.99$; empirical selection is marked by dots.
 Selection hotspots: when sizeable cross-immunity drop on the flank of an antigenic landscape (arrows in a-c) coincides with large immune weights (arrows in d), the fitness model predicts time windows of strong selection for antigenic escape (arrows in e, f mark clade shifts starting in a selection hotspot). 
 Immune weight and fitness trajectories are averaged over regions (see Fig.~S4 for regional trajectories).
 }
\label{fig:antigenic_space}
\end{figure*}

\paragraph*{Fitness trajectories and selection hotspots}
 
The ML fitness model can be applied to the long-term turnover of viral clades up to date, including recent frequency changes between the variants BA.1, BA.2, and BA.4/5 within the $o$ clade (we use shorthands $o1$, $o2$, and $o45$). First, we look at two building blocks of antigenic fitness: antigenic landscapes and immune weights. We 
define antigenic landscapes for each immune channel $k$ by plotting all cross-immunity factors $c_i^k$ against their corresponding titers $T_i^k$ (using a fixed time delay $\Delta t = \tau$ to account for antibody decay in an approximate way; Methods). These landscapes visualise antigenic drift, that is, the partial escape from population immunity by gradual evolutionary steps\cite{Earn2002,Boni2006} (Fig.~4a-c; arrows mark sizeable steps between successive variants). In this picture, the time-dependence of cross-immunity is captured by immune weight functions $Q_k(t)$, which measure recent infections or vaccinations in channel $k$, again over a time window of order $\tau$ (Fig.~4d, Methods). Next, we juxtapose these immune trajectories to long-term trajectories showing the antigenic selection between successive variants, $s_\ag (t)$ (Fig.~4e), and the fitness gap of each variant, $\delta f_i (t) = f_i(t) - \bar f (t)$ (Fig.~4f); these trajectories are computed from equation~(\ref{f}). Fitness gaps are shifted by the mean population fitness $\bar f (t) = \sum_i x_i (t) f_i (t)$ and include the intrinsic component (Methods). As expected from the analysis above, the ML model is in quantitative agreement with empirical selection (dots in Fig.~4f). The trajectories of Fig.~4ef are averaged over 14 regions (for regional trajectories, see Fig.~S4). 

Together, the trajectories of Fig.~4 show a pattern of selection hotspots. The fitness model predicts time windows of strong antigenic selection when antigenic advance on the flank of a Hill landscape generates sizeable cross-immunity loss and coincides with high immune weight. 
We now trace this pattern through successive clade shifts. 
At early stages of evolution, until spring 2021, all immune weights were small (Fig.~4d). Hence, intrinsic selection governed the $\anc - \alpha$ shift; consistently, the antigenic advance (neutralisation titer drop) $\Delta T_{\anc \alpha}^{\vac}$ was small (Fig.~4a). Between spring 2021 and spring 2022, primary and booster vaccination generated the dominant immune weights and induced directional antigenic selection for escape from the vaccine strain, 
while infection-induced immunity remained relatively small (Fig.~4d). The clade shifts $\alpha-\delta$ and $\delta-o$ carried increasing antigenic advance $\Delta T_{\alpha \delta}^{\vac}$ and $\Delta T_{\delta o}^{\vac}$, and smaller advance with respect to boosting (Fig.~4ab). These changes mark the onset of antigenic drift. The fitness model identifies a first clear hotspot of antigenic selection driving the $\delta-o$ shift. At the start of this shift, large cross-immunity change and large immune weight coincided in the primary vaccination channel (orange arrows in Fig.~4ade). This is consistent with the observed dynamics: $\delta-o$ was faster than the previous shifts and under exceptionally high initial selection, $\hat s = 0.15$ (orange arrow in Fig.~4f, Fig. 1b). The following shift, $o1 - o2$, involved subclades with similar neutralisation by primary and booster vaccination. This shift is inferred to be governed predominantly by intrinsic selection (Table~S1 and~S3); consistently, selection is only weakly time-dependent (Fig.~S3).

The most recent viral-immune co-evolution shows two important novelties. In spring 2022, infection immunity increased, while vaccination-induced immunity decreased; both components are reaching comparable weights (Fig.~4d). Whether vaccination remains at sizeable immune weight will depend on availability and acceptance of vaccines in the future. These immune weight changes also mark the onset of immune drift, that is, the response of population immunity to antigenic drift of the viral population. 
Recently, population immunity has shifted its center of mass from wt towards $o$; components cognate to each of these clades have reached comparable weight (red arrows in Fig.~4d). 

At this point, our fitness model predicts the next selection hotspot for novel variants carrying antigenic advance away from vaccination \emph{and} from $o$ infections. The recent antigenic evolution within the $o$ clade follows this scenario: the emerging variants $o4$ and $o5$ (BA.4 and BA.5) combine antigenic advance in three channels\cite{Khan2022b,Hachmann2022} (red arrows in Fig.~4bce). Consistently, these variants show fast initial growth with empirical selection $\hat s = 0.12$ (red arrow in Fig.~4f).
Moreover, near-future mutations carrying antigenic advance in the same channels are predicted to be in the same hotspot (dashed arrows in Fig.~4bc). 

If these emerging variants develop into major clade shifts, 
they will further increase the $o$ immune weight factors. On the other hand, population immunity against earlier variants could be maintained by backboosting\cite{Fonville2014} of $o$ infections or by bivalent vaccines with a wt component.
The future evolutionary trajectories of new variants will also depend on their mutual antigenic relations, which have not yet been assayed comprehensively to date. Together, the recent evolutionary dynamics signals the unfolding of antigenic complexity towards coexistence of multiple antigenic variants and immune classes.

\subsection*{Discussion} 

Here we have established a data-driven, multi-component fitness model for the evolution of SARS-CoV-2. By applying this model to recent evolutionary trajectories in multiple regions, we have quantified intrinsic and antigenic selection driving the genetic and functional evolution of the virus. In particular, primary vaccination impacted on the speed of global clade shifts in 2021. Booster vaccination generated higher cross-protection, but weaker selection for antigenic escape in the same period (Fig.~3). These results underscore that vaccine breadth is important for constraining antigenic escape evolution. More broadly, they highlight the need to integrate evolutionary feedback into vaccine design. 

In the recent evolution of SARS-CoV-2, two general trends are revealed by our analysis. Antigenic selection has increased in strength and has broadened its target:
primary infection by distinct viral variants has generated an increasing number of antigenic selection components (Fig.~3b, Fig.~4). These trends mark the transition from initial, post-zoonotic adaptation of the virus to evolution to an endemic state, where antigenic evolution continues to be fuelled by the buildup of population immunity to circulating viral variants. A plausible end point of this transition becomes clear by comparison with influenza, a long-term endemic virus in humans. In influenza, the viral escape from population immunity follows a specific mode of antigenic drift, where multiple variants with different cross-immunity profiles compete for prevalence\cite{Strelkowa2012}. This mode is marked by continuous, adaptive clade turnover with characteristic time scales of several months, which is substantially slower than the recent prevalence shifts of SARS-CoV-2 variants (Fig.~1a). In contrast, non-antigenic mutations in influenza proteins are under broad negative selection;  observed changes often compensate the deleterious collateral effects of antigenic evolution on conserved molecular traits (including protein stability and receptor binding)\cite{Gong2013,Strelkowa2012,Laessig2017}. If SARS-CoV-2 reaches a similar endemic state, antigenic evolution is expected to slow down and most 
intrinsic changes, e.g., in binding affinity to human receptors, will become compensatory. Recent findings of compensatory evolution leading to Omicron support this scanario\cite{Moulana2022}.

The expected transition of SARS-CoV-2 to gradual, multi-faceted antigenic evolution will open the possibility to predict the future evolution of the viral population by data-driven fitness models\cite{Luksza2014,Morris2018,Huddleston2020} and to inform preemptive vaccination strategies\cite{Mustonen2020}. 
Previous work has established an important prerequisite of predictions: neutralisation assays of human antisera against viral strains quantify the immune protection of human cohorts against secondary infections\cite{Feng2021,Khoury2021}. Here, we have shown that this data can be harvested at the population scale, to compute immune drift and inform antigenic fitness models. As a first step of short-term predictions, we have identified emerging variants in antigenic selection hotspots, in quantitative agreement with their observed clade growth (Fig.~4). This and future predictions of SARS-CoV-2 evolution require integrated analysis of genome sequences, epidemiological records, and increasingly complex antigenic data. While sequence and epidemiological data are already collected in large amounts, our analysis calls for world-wide, real-time tracking of antigenic evolution by cross-neutralisation assays. This will be critical for our ability to predict antigenic escape evolution and to integrate such predictions into vaccine design.

\newpage




\clearpage
\newpage
{\small 

\subsection*{Methods}

\paragraph{\small Sequence data and primary sequence analysis.} The study is based on sequence data from the GISAID EpiCov database\cite{Shu2017} available until 06-22-2022. For quality control, we truncate the 3' and 5' regions of sequences and remove sequences that contain more than 5\% ambigous sites or have an incomplete collection date. We align all sequences against a reference isolate from GenBank\cite{Benson2015} (MN908947), using MAFFT v7.490\cite{Katoh2013}. Then we map sequences to Variants of Concern/Interest (VOCs/VOIs), using the set of identifier amino acid changes given in Outbreak.info\cite{Outbreak}. As a cross-check, we independently infer a maximum-likelihood (ML) strain tree from quality-controlled sequences under the nucleotide substitution model GTR+G of IQTree\cite{Minh2020}, using the reference isolate hCoV-19/Wuhan/Hu-1/2019 (GISAID-Accession: EPI ISL 402125) as root. For assessment of the tree topology, we use the ultrafast bootstrap function\cite{Hoang2017} with 1000 replicates. Internal nodes are timed by TreeTime\cite{Sagulenko2018} with a fixed clock rate of $8 \times 10^{-4}$ under a skyline coalescent tree prior\cite{Kingman1982}. Consistently, variants are mapped to unique genetic clades (subtrees) of the ML tree (Fig.~2b).

\vspace*{-1.5ex} 
\paragraph{\small Frequency trajectories of variants.} 
For a given variant $i$, we define the smoothened count $n_i (t) = Z^{-1} \sum_{\nu \in i} \exp [-(t - t_\nu)^4/\delta^4]$, where the sum runs over all sequences $\nu$ mapped to variant $i$, $t_\nu$ is the collection date of sequence $\nu$, and $Z$ is a normalisation constant. We use a smoothening period $\delta = 33$d. The corresponding variant frequency is then defined by normalisation over all co-existing variants, $x_i (t) = n_i (t) / \sum_i n_i (t)$. These frequency trajectories, evaluated separately for each region of this study, are shown in Fig.~1 and Figs.~S1, S2. 

\vspace*{-1.5ex} 
\paragraph{\small Inference of empirical selection.}
In a population of different variants, the absolute fitness of each variant is defined as the growth rate of its population, 
\EQ
f_i (t) = \frac{\dot N_i (t)}{N_i(t)}
\EE
($i = 1, \dots, n$). The absolute fitness is related to its reproductive number, defined as the mean number of new infections generated by an individual during its infectious period $\tau_0$,
\EQ
R_i (t) = \exp[\tau_0 f_i (t)]. 
\label{R} 
\EE
The fitness difference (selection coefficient) between a given pair of variants, $s_{ij} (t) = f_i (t) - f_j (t)$, is given by $s_{ij} (t) = ({\rm d}/{\rm dt})(\log N_i (t) - \log N_j(t)) = ({\rm d}/{\rm dt}) \log (( N_i (t) /  N_j(t))$, independently of the other co-circulating variants. This relation can also be written in terms of the population frequencies $x_i (t) = N_i (t) / \sum_k N_k (t)$, leading to equation (\ref{eq:s_hat}) of the main text. 

For each clade shift and each region included in the study, we infer a trajectory of empirical selection, 
\EQ
\hat \s = (\hat s(t_1), \hat s(t_2), \dots, \hat s (t_n)), 
\label{straj} 
\EE
which records the time-dependent fitness difference between invading and ancestral strain, $\hat s (t_i) = f_{\rm inv} (t_i) - f_{\rm anc} (t_i)$ ($i = 1, \dots, n$). A hat distinguishes these empirical selection coefficients from their model-based counterparts introduced below. At each point of the trajectory, we evaluate the selection gradient of equation~(\ref{eq:s_hat}), 
\EQ
\hat{s}(t_i) = \frac{1}{\Delta t} \left[ \log \left ( \frac{x_{\rm inv} (t_i + \Delta t/2)}{x_{\rm anc} (t_i + \Delta t/2)} \right ) 
- \log \left ( \frac{x_{\rm inv} (t_i - \Delta t/2)}{x_{\rm anc} (t_i - \Delta t/2)} \right ) \right] 
\qquad (i = 1, \dots, n), 
\label{sint}
\EE
using a time window $\Delta t = 30$d for the $\anc-\alpha$ and $\delta - o$ shifts and $\Delta t = 40$d for the $\alpha - \delta$ shift (which extends over a longer period). Increasing $\Delta t$ reduces the statistical error of $\hat{s}(t_i)$ but reduces the time span covered by a trajectory $\mathbf{\hat{s}}$.}  We evaluate equation~(\ref{straj}) for the maximal time interval such that $x_{\rm anc} (t_i \pm \Delta t/2) > 0.01$ and $x_{\rm inv} (t_i \pm \Delta t/2) > 0.01$ along the entire trajectory. The start point $t_1$ is the first day when $x_{\rm inv} (t - \Delta t/2) > 0.01$. From this point, selection is recorded weekly, $t_i - t_{i-1} = 7$d ($i = 2, \dots, n$),  and $t_n$ is the last point of this sequence where $x_{\rm anc} (t + \Delta t/2) > 0.01$. Single measurements $\hat{s}(t_i)$ are excluded when at least one of the sequence counts $n_{\rm anc}(t_i \pm \Delta t/2)$ or $n_{\rm inv}(t_i \pm \Delta t/2)$ is $<10$. Statistical errors for selection trajectories are evaluated by binomial sampling of counts $n_{\rm anc} (t)$ and $n_{\rm inv} (t)$ with a pseudocount of 1. Empirical selection trajectories are reported in Fig.~1b and Figs.~S1-S3. 

For the subsequent analysis, we grade the complete clade shifts $\anc - \alpha$, $\alpha - \delta$, $\delta - o1$, $o1 - o2$ by the time dependence of their empirical selection trajectories (Fig.~S3). We evaluate two summary statistics: (i)~the amount of systematic time-dependent variation of selection, defined as $\mbox{Var} (s_{\rm lin})$, averaged over regions, where $s_{\rm lin} (t)$ is a linear regression to the ensemble of trajectories; (ii)~the statistical significance of the linear regression, $P$ (two-sided Wald test). This identifies two shifts with substantial, statistically significant time-dependent variation of selection, $\alpha - \delta$ and $\delta - o$.

\vspace*{-1.5ex} 
\paragraph{\small Infection and vaccination trajectories. } 
Daily vaccination and infection rates for individual regions have been obtained from Ourworldindata.org\cite{owidcoronavirus} and from CDC COVID Data Tracker\cite{coronaUSA} for US states (download date: 06-22-2022). Clade-specific infection rates $y_k (t)$ are computed by multiplying the total daily infection rates reported in each region with the simultaneous viral clade frequencies $x_k (t)$. The resulting cumulative population fractions of infected individuals, $y_k (t)$, together with cumulative population fractions of primary and booster vaccinations, $y_\vac (t)$ and $y_\bst (t)$, are reported in Figs.~S1-S2.

\vspace*{-1.5ex} 
\paragraph{\small Data integration for regional analysis.} 
This study is based on sequence data and epidemiological data from multiple regions (countries and US states) for parallel analysis. Sequence data is used to infer empirical selection trajectories for individual clade shifts, as defined in equations~(\ref{straj}) and~(\ref{sint}). Epidemiological records provide input to the antigenic fitness model, equations~(\ref{C}) and~(\ref{f}). Evaluation of the fitness model, which is detailed below, integrates data of both categories and requires stringent criteria of data availability and comparability. 

To enable this analysis, we choose the set of countries to be included in model inference based on uniform criteria. Additionally, we include 3 US states (New York, Texas, California), each representative of a different geographic region, that satisfy the same criteria. For each clade shift ${\rm anc} - {\rm inv}$, we require the following: 
(i)~anc and inv are majority variants at times $t$ and $t' > t$ of the clade shift, respectively; i.e., $x_{\rm anc} (t) > 0.5$ and $x_{\rm inv}(t) > 0.5$. This criterion excludes regions where other variants are prevalent during the shift ${\rm anc} - {\rm inv}$ (e.g., Brazil and South Africa have $x_{\alpha} < 0.5$ throughout the $\anc - \alpha$ shift). 
(ii)~anc and inv have a combined, smoothened sequence count $n_{\rm anc} (t) + n_{\rm inv} (t) > n_0$ throughout the clade shift. This criterion ensures that the empirical frequencies $x_{\rm anc} (t)$ and $x_{\rm inv} (t)$, especially minority frequencies, can be estimated with reasonable statistical errors. We use threshold values $n_0 = 500$ for $\anc-\alpha$ and $\alpha - \delta$ and $n_0 = 750$ for $\delta - o$ (reflecting the increased sequence availability).
(iii)~The empirical selection trajectory $\hat{\mathbf{s}}$ contains at least 4 ($\anc - \alpha$, $\delta - o$) or 6 ($\alpha - \delta$) measured points $\hat{s}(t_i)$; the threshold values reflect the relative duration of shifts. This criterion ensures a sufficient signal-to-noise ratio for inference of temporal variation along the trajectory. 
(iv) In the $\delta - o$ shift, the cumulative fraction of $o$ infections exceeds a threshold value, $y_{o} > 0.01$. The $o$ variant, which is characterised by many less severe cases, is likely to be particularly affected by underreporting. This criterion excludes regions with very low $o$ count ($y_o$ is less than $\sim$ 20\% of the remaining regions) and ensures that cross-immunity trajectories, as given by equation~(\ref{C}), can be evaluated across regions with sufficient consistency. (v)~Vaccinations have been predominantly by mRNA vaccines and epidemiological records in the database\cite{owidcoronavirus} are complete. This criterion ensures that  antigenic data for mRNA vaccines can be used uniformly (Table~S1). It excludes regions with substantial use of viral vector vaccines (e.g., the UK) and with partial records (e.g., for booster vaccinations in Sweden and Croatia). 

Based on these criteria, our analysis includes (i)~11 regions for the $\anc-\alpha$ shift, (ii)~16 regions for the $\alpha - \delta$ shift, and (iii)~14 regions for the $\delta - o$ shift (Fig.~1, Fig.~3, Figs. S1-S3). Regions analysed for both $\alpha - \delta$ and $\delta -o$ are used for the long-term trajectories (Fig.~4, Fig.~S4). Scope and limitations of this set of regions for the inference of selection are described below.

\vspace*{-1.5ex} 
\paragraph{\small Antigenic data.}
Neutralisation assays for SARS-CoV-2 test the potency of antisera induced by a given primary immunisation to neutralise viruses of different variants. Log dilution titers measure the minimum antiserum concentration required for neutralisation, 
\EQ
T_i^k = \log_2 \frac{K_0}{K_i^k}, 
\EE
relative to a reference concentration $K_0$. 
Hence, $\log_2$ titer differences, or neutralisation fold changes, $\Delta T_{ij}^k  \equiv \Delta T_i^k  - T_j^k$, measure differences in antigenicity between variants, $\Delta T_{ij}^k  = \log_2 (K_j^k/K_i^k)$. 
 We note that these differences 
 are specific to each primary challenge (immune channel) $k$. For example, the inequality $T_{\alpha \delta}^\bst < T_{\alpha \delta}^\vac$ reflects the increased breadth of booster vaccinations compared to primary vaccinations. In contrast, uni-valued antigenic distances between variants, $d_{ij}$, can be computed from the titer matrix $(T_i^k)$ by multi-dimensional scaling methods\cite{Smith1997,Smith2004}. Such distance measures average over inhomogeneities between immune channels.

Here we define a matrix of titer drops $\Delta T_{i}^k$, 
\EQ
\Delta T_{i}^k = T_*^k - T_i^k 
\qquad \mbox{($i = \alpha$, $\delta$, $o$ ($o1$), $o2$, $o45$; \; $k = \alpha$, $\delta$, $o$ ($o1$), $o2$, $o45$, vac, bst)}, 
\label{drop} 
\EE 
with respect to a reference for each immune channel, $T_*^k = T_k^k$ ($k = \alpha, \delta, o(o1), o2, o45$) and $T_*^k = T_\anc^k$ ($k = \vac, \bst$). This procedure eliminates technical differences between assays in absolute antibody concentration. We assemble this matrix in Table~S1, using primary data from different sources\cite{Bates2021,Cameroni2021,Garcia2021,Liu2021c,Zhou2021b,Wang2021c,Liu2021b,Mlcochova2021,Planas2021,Planas2021b,Muik2021,Uriu2021,Roessler2022,Gruell2022,Planas2021c,Straten2022,Wilks2022,Iketani2022,Bowen2022,Cao2022,Mykytyn2022,Hachmann2022,Wang2022c,Khan2022b}. 
We proceed as follows: 
(i)~For matrix elements with available data, $\Delta T_i^k$ is the average of the corresponding primary measurements. This procedure eliminates technical differences between assays in absolute antibody concentration. As appropriate for the analysis in our set of regions, all vaccination titers refer to mRNA vaccines.
(ii)~If no data are available for $\Delta T_i^k$ but the conjugate titer $\Delta T_k^i$ has been measured, we use the approximate substitution $\Delta T_i^k \approx \Delta T_k^i$, as discussed in ref.~[\citen{Neher2016}]. 
(iii)~If no data are available for $\Delta T_i^k$ but the titer $\Delta T_j^k$ of a closely related clade has been measured, we use the approximate substitution $\Delta T_i^k \approx \Delta T_j^k$, which should be understood as a lower bound (this applies to the recent variants $o2$ and $o45$). 

The matrix of absolute neutralisation titers, $T_i^k$, is then computed by equation~(\ref{drop}), combining the titer drops $\Delta T_i^k$ of Table~S1 and the reference titers $T_*^k = 6.5$, ($k = \alpha, \delta, o(o1), o2, o45$), $T_*^\vac = 7.8$, $T_*^\bst = 9.8$ reported in ref.~[\citen{Polack2020}]. A titer difference between vaccination and booster, $T_*^\bst - T_*^\vac \approx 2.0$, has been observed in several studies\cite{Planas2021c,Gruell2022,GarciaBeltran2022}. The titers $T_i^k$ enter the cross-immunity functions $c_i^k$, $C_i^k (t)$, and $\bar c_i^k$ defined below and are shown in Fig.~2ab.

The decay of antibody concentration after primary immunisation has been characterised in recent work\cite{Iyer2020,Israel2022}. Here we describe this effect by a linear titer reduction with time after primary challenge, 
\EQ
T_i^k (\Delta t) = T_i^k - \frac{\Delta t}{\tau}, 
\label{DeltaT}
\EE 
corresponding to an exponential decay of antibody concentration, with a uniform decay time $90$d (i.e., half life $\tau = 65$d). This is broadly consistent with experimental data; we infer decay times in the range $[60, 170]$d from several studies\cite{Iyer2020,Israel2022,Planas2021b,Planas2021c,Gruell2022}. In addition, we check that varying $\tau$ in this range does not affect our results (in particular,  the rank order of variants with respect to antigenic fitness remains unchanged).

\vspace*{-1.5ex} 
\paragraph{\small Cross-immunity trajectories.}
The cross-immunity factor $c_i^k$ is defined as the relative reduction in infections by variant $i$ induced by (recent) immunisation in channel $k$. 
As shown in recent work\cite{Khoury2021,Feng2021}, absolute titers of SARS-CoV-2 neutralisation assays can predict cross-immunity, $c_i^k = H(T_i^k)$ with 
\EQ 
H (T) =  \frac{1}{1 + \exp[- \lambda (T - T_{50})]}.
\label{c} 
\EE
This relation has been established in ref.~[\citen{Khoury2021}] with constants $T_{50} = 4.2$ and $\lambda = 0.9$. The resulting cross-immunity factors $c_i^k (\Delta t)$ include antibody decay, as given by equation~(\ref{DeltaT}). Hence, they depend on the time since primary immunisation, 
\EQ
c_i^k (\Delta t) = H \big (\mbox{$T_i^k - \frac{\Delta t}{\tau}$} \big ). 
\label{cDt} 
\EE
These factors enter the population cross-immunity functions $C_i^k (t)$, equation~(\ref{C}), which become 
\EQ
C_i^k (t) = \int^t H \big (T_i^k - \mbox{$\frac{t - t'}{\tau}$} \big ) \, \dot y(t) \, dt'.
\label{C2} 
\EE
These functions enter all evaluations of the fitness model, equation~(\ref{f}) (Fig. 1c, Fig.~3, Fig.~4ef, Figs. S1, S2, and S4). 

To display the emergence of selection hotspots, we approximate the cross-immunity functions, equation~(\ref{C2}), by time-independent effective factors and immune weights. (i)~The effective cross-immunity factors $\bar c_i^k$ are obtained from equation~(\ref{cDt}) at a fixed time delay $\Delta t = \tau$ after primary immunisation, $\bar c_i^k = H (T_i^k - 1)$, 
which accounts for the decay of immune response in an approximate way. These factors define the antigenic landscapes shown in Fig.~4a-c. 
(ii)~The immune weight functions, 
\EQ
Q_k (t)  =   \int_{- \infty}^t  H \big (T_0 - \mbox{$\frac{t - t'}{\tau}$} \big ) \, \dot y (t') \, dt',
\label{Qk} 
\EE
measure the effective population fractions of immune individuals in channel $k$. They account for immune decay from a fixed reference titer $T_0 = 6.5$ and follow the time-dependence of cross-immunity functions $C_i^k (t)$ and selection coefficients $s_k (t)$ in an approximate way (Fig.~4de, Fig.~S4). Selection hotspots emerge if large steps on an antigenic landscape, $\bar c_{\rm inv}^k - \bar c_{\rm anc}^k$, coincide with sizeable immune weights $Q_k (t)$ in one or more immune channels (Fig.~4).

\vspace*{-1.5ex} 
\paragraph{\small Fitness model.} 
Equation~(\ref{f}) expresses the fitness of viral variants as a sum of intrinsic and antigenic fitness components, $f_i (t) = f^0_i + \sum_k f_i^k (t) $. Intrinsic fitness, $f_0$, integrates contributions from several molecular phenotypes, including protein stability, host receptor binding, and traits related to intra-cellular viral replication. The antigenic components $f_i^k (t)$ describe the impact of antibody binding on viral growth, summed over the immune repertoire components of different channels of primary infection or vaccination. Importantly, the input of this fitness model can be learned by integration of sequence data, epidemiological records, and antigenic assays. 

The additive form the fitness model neglects epistasis between fitness components. The additivity assumption is justified between intrinsic and antigenic fitness, because these components are associated to different stages of the viral replication cycle. The additivity of antigenic fitness components rests on the approximation of a well-mixed host population and short infection times. In this approximation, each viral lineage is subject to a dense sequence of random encounters with  hosts of different immune channels $k$, leading to averaging of antigenic fitness effects. Multiple infections in an individual can generate additional immune channels; however, these effects are relatively small over the short periods of SARS-CoV-2 evolution studied in this paper.

Our analysis of the fitness model focuses on selection coefficients between co-existing variants, 
\EQ
s_{ij} (t) \equiv f_i(t) - f_j (t) = s_{ij}^0 - \sum_k \gamma_k \big [ C_i^k (t) - C_j^k (t) \big ], 
\EE
because these can directly be compared with their empirical counterparts $\hat s_{ij} (t)$. Of equal importance, selection coefficients within a region decouple from the changes in viral ecology within that region.
Specifically, seasonality and contact limitations can generate strongly time-dependent reproductive numbers. However, any modulation of the form $R (t) \to \alpha (t) R(t)$ leaves the selection coefficients $s_{ij}$ invariant, as can be seen from equation~(\ref{R}). Our inference of empirical selection, as described above, is also independent of the underlying infectious period $\tau_0$, which may itself be under evolutionary pressure and change with time\cite{Hart2022a,Hart2022b}. To keep this independence, we report all selection coefficients in fixed units [1/d].

\vspace*{-1.5ex} 
\paragraph{\small Inference of fitness model parameters.}
The free parameters $\gamma_k$ ($k = 1, \dots, n$) measure the fitness effect of each cross-immunity component. These parameters calibrate the model to data of complex real populations differing, for example, in population structure (including incidence structure), infection histories, and monitoring of infections. To avoid overfitting, we use a minimal model with just 3 global antigenic parameters: (i) A basic rate $\gamma_\vac = \gamma_\bst$ translates cross-immunity generated by vaccination into units of selection. 
(ii)~This rate is downweighted to a value $\gamma_\vac' = a \gamma_\vac$ for the shift $\delta - o$ and later shifts. This can be seen as a heuristic to account for the effect of double infections\cite{Bates2022}, which increase cross-immunity and decrease cross-immunity differences between variants. 
(iii)~Cross-immunity in all infection channels is uniformly upweighted,  $\gamma_k = b \gamma_\vac$, to account for underreporting of infections relative to vaccinations.  

Our inference proceeds in two steps. First, we train the antigenic fitness model using data from the clade shifts $\alpha - \delta$ and $\delta - o$. These shifts are suitable because they carry sizeable antigenic advance $\Delta T_{\alpha \delta}^\vac$ and $\Delta T_{\delta o}^\vac$ (Fig.~2a), selection shows a substantial and statistically significant time dependence (Fig.~1b), and population immunity has started to pick up (Fig.~4d). We infer the ML likelihood model by aggregation of log likelihood scores over the sets of regional selection trajectories for the clade shifts $\alpha - \delta$ and $\delta - o$. We use the score function 
\EQ
 L(\hat \s, \s) = - \sum_{i=1}^n \frac{(\Delta \hat s(t_i) - \Delta s (t_i))^2}{2 \sigma^2 (t_i)} 
 \label{Ls} 
\EE
for a single empirical selection trajectory $\hat \s$, equation~(\ref{straj}), and its model-based counterpart $\s$. This score evaluates selection change, $\Delta s (t) = s(t) - \langle s \rangle$, where brackets denote averaging over time. Hence, the fitness model is trained on the time-dependence of selection in each region, in order to avoid the confounding factor of heterogeneity across regions. The expected square deviation is $\sigma^2 (t_i) = \sigma^2_s (t_i) + \sigma_0^2$; the first term describes the sampling error of sequence counts, which enters frequency and empirical selection estimates, the second term summarises fluctuations unrelated to sequence counts. The total log likelihood score is the sum $L =  \sum L(\s, \hat \s)$, which runs over both shifts and all included regions. Table~S2 lists the ML parameters $\gamma_\vac, a, b$ and the ML score $L$ relative to a null model of time-independent selection (see below). The 95\% confidence intervals of the inferred parameters are computed by resampling the empirical selection data with fluctuations $\sigma^2$.
We note that the ML values $a < 1, b > 1$ are consistent with the interpretation of these parameters as weighting factors accounting for double-infections and underreporting (see above). Second, we infer the intrinsic selection for each shift as the difference between empirical selection and ML antigenic selection, $s_0 = \langle \! \langle \hat s - s_\ag \rangle \! \rangle$, where the double brackets denote averaging over time and regions. The ML antigenic selection coefficients, $\langle \! \langle s_k \rangle \! \rangle$, and the intrinsic selection coefficient $s_0$ between invading and ancestral variant are listed 
in Table~S3; see also Fig.~3b. Confidence intervals are computed by resampling model parameters with their confidence intervals. Consistently, we infer weak antigenic selection for the shifts $\anc - \alpha$ and $o1 - o2$, which also show only weak time dependence of selection (Fig.~S3).

\vspace*{-1.5ex} 
\paragraph{\small Significance analysis of the fitness model.} 
To assess the statistical significance of our inference, we compare four fitness models of the form (\ref{f}): the full model used in the main text (VI: antigenic selection by vaccination and infection, intrinsic selection), two partial models (V: antigenic selection only by vaccination, intrinsic selection;  I: antigenic selection only by infection, intrinsic selection), and a null model (0: intrinsic selection only). We infer conditional ML parameters for each model and we rank models by their ML score difference to the null model, $\Delta L = L - L_0$ (Table~S2). An alternative ranking by BIC score\cite{Schwarz1978}, which contains a score penalty for the number of model parameter, leads to the same result. We observe the following: 
(i)~All antigenic fitness models have significantly higher scores than the null model, which shows that the empirical selection data are incompatible with time-independent selection. 
(ii)~The full model has a significantly higher score than any of the other models; both vaccination and infection are significant components of antigenic selection. 
(iii)~Vaccination explains a larger part of the time-dependent data than infection ($\Delta L_V > \Delta L_I$), which is consistent with the ranking of selection coefficients inferred from the full model (Fig.~3b, Table~S3). 
(iv)~The score gain of the full model is less than the sum of its parts ($\Delta L_{IV} < \Delta L_V + \Delta L_I$).  This can be associated with statistical correlations in the input data for both antigenic model components. For example, the fraction of vaccinated individuals $y_{\rm vac}$ is weakly anti-correlated with the fraction of $\delta$ infections, $y_{\delta}$.

\vspace*{-1.5ex} 
\paragraph{\small Fitness trajectories.} 
Long-term fitness trajectories display clade turnover of multiple successive shifts (Fig.~4f, Fig.~S4). For each of the variants $\alpha, \delta, o (o1), o2, o45$, we plot the time-dependent fitness gap, $\delta f_i (t) = f_i (t) - \bar f (t)$, where $\bar f (t) = \sum_j x_j (t) f_j (t)$ is the mean population fitness. Like selection coefficients, fitness gaps decouple from ecological factors affecting absolute growth (see the discussion above). Assuming that the fitness difference between ancestral and invading variant, $s(t)$, is dominant during each crossover, we obtain the fitness gap trajectories $\delta f_{\rm anc} (t) = - s(t) x_{\rm inv} (t)$ and $\delta f_{\rm inv} (t) = - s(t) [1 - x_{\rm inv} (t)]$, as well as their empirical counterparts $\delta \hat f_{\rm anc} (t)$ and $\delta \hat f_{\rm inv} (t)$. For each variant, we patch trajectories from origination to near-fixation (here in the time interval $(t_{0,i}, t_{f,i})$ given by $x_i (t_{0,i}) = 0.01$ and $x_i (t_{f,i}) = 0.99$). The long-term trajectories display selection hotspots and confirm the quantitative agreement between empirical and model-based fitness.

\vspace*{-1.5ex} 
\paragraph{\small Model-based inference of selection across regions.} 
 As shown by the preceding analysis, we can infer a statistically significant fitness model with few, global parameters from sequence and epidemiological data aggregated over a set of regions and combined with antigenic data. The model describes common time-dependent patterns of selection in these regions and serves two main purposes: to provide a breakdown of selection in intrinsic and antigenic components (Fig.~3) and to display selection hotspots in long-term trajectories (Fig.~4). Our inference procedure rests on stringent criteria for the joint availability of sequence and epidemiological data in each of these regions (as listed above). A number of points support this procedure: 
(i)~The results are robust under variation of the inclusion criteria for regions. In particular, the signal of antigenic selection in data and model is broadly distributed over regions (Figs.~S1-S3). Hence, the selection averages reported in Fig.~3b and Table~S3 are reproducible in subsampled sets of regions. 
(ii)~Within the set of regions included, the model is applicable beyond the $\alpha-\delta$ and $\delta - o$ shifts used for training. The early $\anc - \alpha$ shift and the recent $o1-o2$ shifts serve as controls. In both cases, we infer weak antigenic selection, consistent with weak time dependence of empirical selection (Fig.~S4). For the emerging $o2 - o45$ shift, strong antigenic selection is consistent with fast initial growth of the new variants. 

Our model-based inference of selection excludes a number of regions that do not fulfil the criteria of joint data availability. 
 (i)~For the $\alpha-\delta$ shift, several regions are excluded because VOCs other than $\alpha$ were majority variants prior to the shift to $\delta$ (for example, Beta in South Africa and Gamma in Brazil). Unlike $\alpha - \delta$, these shifts do not involve antigenic advance in the vaccination channel; i.e., $\Delta T^\vac_{{\rm anc} \, \delta} < 0$. However, we lack comprehensive antigenic data on other VOCs as input for the fitness model. 
(ii)~For the $\delta - o$ shift, several regions are excluded because of low reported incidence (e.g., Brazil, California, India, Mexico, Poland, South Korea, Turkey, Texas). Most of these regions show a signal of time-dependent selection consistent with the regions included; however, much lower reported incidence counts prevent reliable immune tracking of infection channels during the $\delta-o$ shift. Variation in reported incidence can, in principle, be incorporated into the fitness model by region-dependent $\gamma_\vac$ factors, but this would likely lead to overfitting. We conclude that at current levels of data availability, a comprehensive cross-regional analysis is not feasible.

\vspace*{-1.5ex} 
\paragraph{Model-free inference of selection in regions with low vaccination coverage.}
 Countries with low vaccination coverage during the $\alpha - \delta$ shift ($y_\vac < 0.1$) disqualify for the model-based analysis because $\alpha$ was not a majority variant (India, Malaysia, Russia, Philippines, Indonesia, South Africa, South Korea) or sequence counts are too low for the inference of selection trajectories (Australia). For this set of countries, we can still infer a selection coefficient $s$ by fitting a sigmoid function to the frequency trajectory $x_\delta (t)$ (to be interpreted as the growth difference between $\delta$ and the average of all other coexisting variants). We find lower region-averaged selection in the set of low-vaccination countries compared to other countries ($ \langle \! \langle \hat s \rangle \! \rangle = 0.08$ vs.~$ \langle \! \langle \hat s \rangle \! \rangle = 0.12$). This is qualitatively consistent with our model-based inference of vaccination-induced selection; however, the genetic heterogeneity of this clade shift prevents a systematic breakdown of antigenic selection into immune channels.

\vspace*{-1.5ex} 
\paragraph{\small Data availability. } 
The datasets analysed in this study are available in published work.

\paragraph{\small Code availability. } 
 The code used in this study is available at \url{https://github.com/m-meijers/vaccine_effect}

\vspace*{-1.5ex} 
\paragraph{Acknowledgements.}       
We thank Florian Klein and Kanika Vanshylla for discussions. This work has been partially funded by Deutsche Forschungsgemeinschaft grant CRC 1310 {\em Predictability in Evolution}.

\newpage

\subsection*{Supplementary Tables and Figures}

\begin{table}[h!]
{\footnotesize {\bf Table S1: Antigenic data. } 
\begin{footnotesize}
\begin{center} 
\begin{tabular}{l | c | c | c | c | c | c | c}
 & $\alpha$ & $\delta$ & $o$ ($o1$) & $o2$ &  $o45$ & vac & bst 
\\
\hline
$\alpha$ & 0 & 1.8 & \emph{5.0} & \emph{5.0} & \emph{5.0} & 0.8 & \emph{0.8} 
\\ 
$\delta$  & 1.5 & 0 & \emph{4.8} & \emph{4.8} & \emph{4.8} & 1.7 & 1.5
\\
$o$ ($o1$) & 5.0 & 4.8 & 0 & 2.1 &              \emph{2.2} & 5.6 & 2.7
\\ 
$o2$ & $< \! \emph{5.0}$ & $< \!$  \emph{4.8} & 1.4 & 0 &      \emph{1.2}  & $< \!$  \emph{5.6} & 2.5 
\\
$o45$ & $< \!$  \emph{5.0} & $< \!$  \emph{4.8} & 2.2 & 1.2 &    0         & $< \!$  \emph{5.6} & 3.9 
\end{tabular}
\end{center}

\end{footnotesize}
\caption*{\footnotesize We list log titer drops, or neutralisation fold changes, $\Delta T_{i}^k = T_*^k - T_i^k$, of strains from variant $i$ assayed against human antisera induced by primary immunisation (infection or vaccination) with strains of variant $k$ (columns). Numbers are average values of primary data from 
ref.~[\citen{Bates2021,Cameroni2021,Garcia2021,Liu2021c,Zhou2021b,Wang2021c,Liu2021b,Mlcochova2021,Planas2021,Planas2021b,Muik2021,Uriu2021,Roessler2022,Gruell2022,Planas2021c,Straten2022,Wilks2022,Iketani2022,Bowen2022,Cao2022,Mykytyn2022,Hachmann2022,Wang2022c,Khan2022b}]. All vaccination titers refer to mRNA vaccines. Where no primary data is available, titer drops are inferred by symmetry or (as lower bounds) by genetic similarity  (numbers in italics, Methods). Absolute titers $T_i^k$ are shifted by the reference titers $T_*^k = 6.5$, ($k = \alpha, \delta, o(o1), o2, o45$), $T_*^\vac = 7.8$, $T_*^\bst = 9.8$ obtained from ref.~[\citen{Polack2020,Planas2021c,Gruell2022,GarciaBeltran2022}]; see Methods and Fig.~2a. 
}
 }
\end{table} 
\bigskip \bigskip

\begin{table}[h!]
\footnotesize {\bf Table S2: Ranking of fitness models. } 
\begin{footnotesize}
\begin{center} 
\begin{tabular}{ c | c | c | c | c | c }
model & \multicolumn{3}{c|}{antigenic parameters}    & \multicolumn{2}{|c}{posterior scores}   
\\
\hline 
 & $ \hs \gamma_\vac \hs $ & \hs $a$ \hs & \hs $b$  \hs & \hs $\Delta L \hs $ & \hs $\Delta H$ \;\;
\\ \hline
VI & $1.22 \pm 0.03$ & $0.24 \pm 0.03$ & $2.0 \pm 0.5$ & 947 & -1883 
\\ \hline
V & 1.22 & 0.34 & - &  882 & -1758 
\\ \hline
I & 4.9 & 0.21 & - &  378 & -750
\\ \hline
0 & - & - & - &  0 & 0 
 
\end{tabular}
\end{center}
\end{footnotesize}
\caption*{\footnotesize 
We compare the full fitness model used in the main text (VI: vaccination + infection + intrinsic selection) with partial models (V: vaccination + intrinsic selection, I: infection + intrinsic selection) and a null model (0: intrinsic selection only). Columns from left to right:
model parameters,
$\gamma_\vac$, $a$, $b$, ML values and 95\% confidence intervals (definitions are given in Methods); 
log likelihood difference to the null model, $\Delta L$; 
BIC score difference to the null model, $\Delta H$. 
}
\end{table} 
\bigskip \bigskip

\begin{table}[h!]
\footnotesize {\bf Table S3: Intrinsic and antigenic selection components.}
\begin{scriptsize}
\begin{center} 
\begin{tabular}{c | c | c | c | c | c | c | c | c}
clade shift   & \multicolumn{8}{|c}{selection coefficients}  
\\
\hline 
  & ${\alpha}$ & ${\delta}$ & ${o \, (o1)}$ & ${o2}$ & ${\rm vac}$ & ${\rm bst}$ &  0 & $s$ 
\\ \hline
$\anc - \alpha$ & $<$ & - & - & - & $<$ & - & $.08 \pm .001$ & $.08 \pm .01$
\\ \hline
$\alpha - \delta$ & $.01 \pm .001$ & $<$ & - & - & $.04 \pm .002$ & - & $.05 \pm .002$ & $.09 \pm .02$
\\ \hline
$\delta - o$ & $<$ & $.02 \pm .006$ & $-.01 \pm .002$ & - & $.06 \pm .01$ & $-.01 \pm .002$ & $.06 \pm .01$ & $.14 \pm .03$
\\ \hline
$o1 - o2$ & $<$ & $<$& $.01 \pm .004$ & $<$ & $<$ & $<$ & $.08 \pm .002$ & $.08 \pm .01$
\\ \hline
$o2 - o45$ & $<$ & $<$ & $.01 \pm .002$ & $.01 \pm .002$ & $<$ & $.04 \pm .01$ & $.06 \pm .008$ & $.12\pm .02$
\end{tabular}
\end{center}
\end{scriptsize}
\caption*{\footnotesize 
Selection coefficients between the invading and the ancestral variant, $s = f_{\rm inv} - f_{\rm anc}$, and their decomposition into antigenic and intrinsic components are inferred for the full fitness model; all values are time averages for each clade shift. 
Rows from top to bottom: major clade shifts, $\anc - \alpha$, $\alpha - \delta$, $\delta - o$; recent clade shifts, $o1 -o2$, $o2 - o45$ (shift incomplete, entries refer to initial period).
Columns from left to right: average antigenic selection in immune channels $k = \alpha$, $\delta$, $o$ ($o1$), $o2$, $\vac$, $\bst$; intrinsic selection (0); total selection ($s$). Selection coefficients are given in units [1/day]; the symbol ``$<$'' marks values $s < 0.01$. We list ML values with 95\% confidence intervals (for selection components) or with rms cross-region variation of selection (for $s$; cf.~Fig.~1b). 
}
\end{table}

\setcounter{figure}{0}
\captionsetup[figure]{labelfont={bf},name={Fig.S},labelsep=space}
\renewcommand{\thetable}{S\arabic{table}}

\newpage
\begin{figure*}[t]
\begin{center}
\includegraphics[width= 0.78 \linewidth]{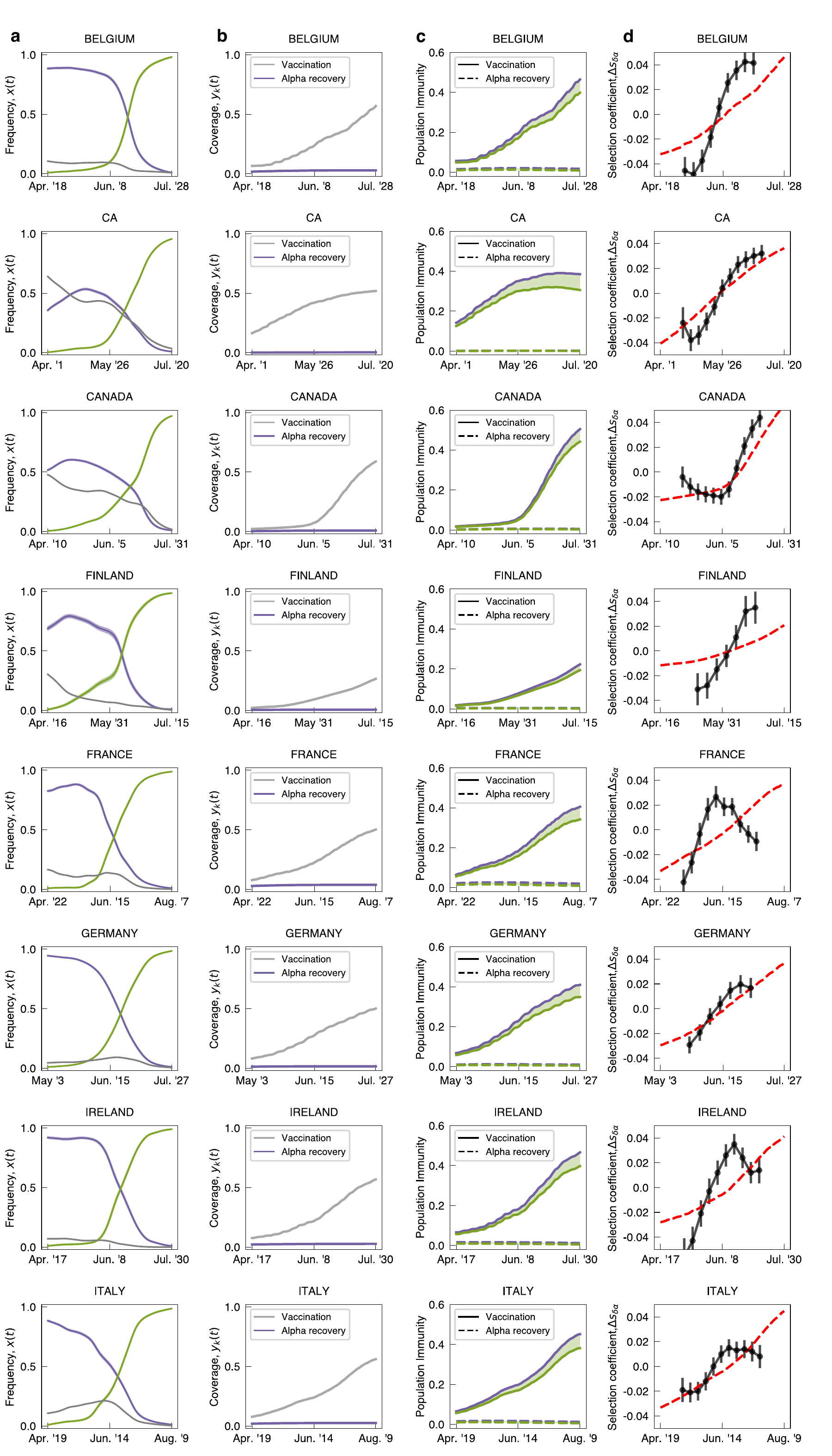}
\end{center}
\begin{flushright} 
\vspace*{-0.5cm}
{\footnotesize (continued on next page)}
\vspace*{-0.5cm}
\end{flushright}
\noindent 
\vspace*{-0.5cm}
\caption*{\footnotesize {\bf Fig. S1}:
{\bf Empirical and model-based trajectories of the $\alpha - \delta$ shift.}  Evolutionary, epidemiological, and cross-immune trajectories are shown for all regions of this study. 
 {\bf (a)}~Observed frequency trajectories of relevant clades, $x_i (t)$; rms sampling error is indicated by shading.
 {\bf (b)}~Cumulative coverage of primary vaccination, $y_\vac (t)$ (light gray), and of booster vaccination, $y_\bst (t)$ (dark gray); cumulative population fraction of $\alpha$ infections, $y_\alpha (t)$ (purple), and of $\delta$ infections, $y_\delta (t)$ (green). 
  {\bf (c)}~Population immunity functions, $C_i^k (t)$ (as in Fig.~1c). 
 {\bf (d)}~Empirical selection change, $\Delta \hat s (t)$ (dots, with rms statistical errors indicated by bars), together with ML model prediction, $\Delta s (t)$ (dashed line). Criteria for inclusion of regions are given in Methods.
 }
 \label{fig:S2}
\end{figure*}

\newpage
\begin{figure*}[t!]
\begin{center}
\vspace*{-1.3cm} 
\includegraphics[width= 0.78 \linewidth]{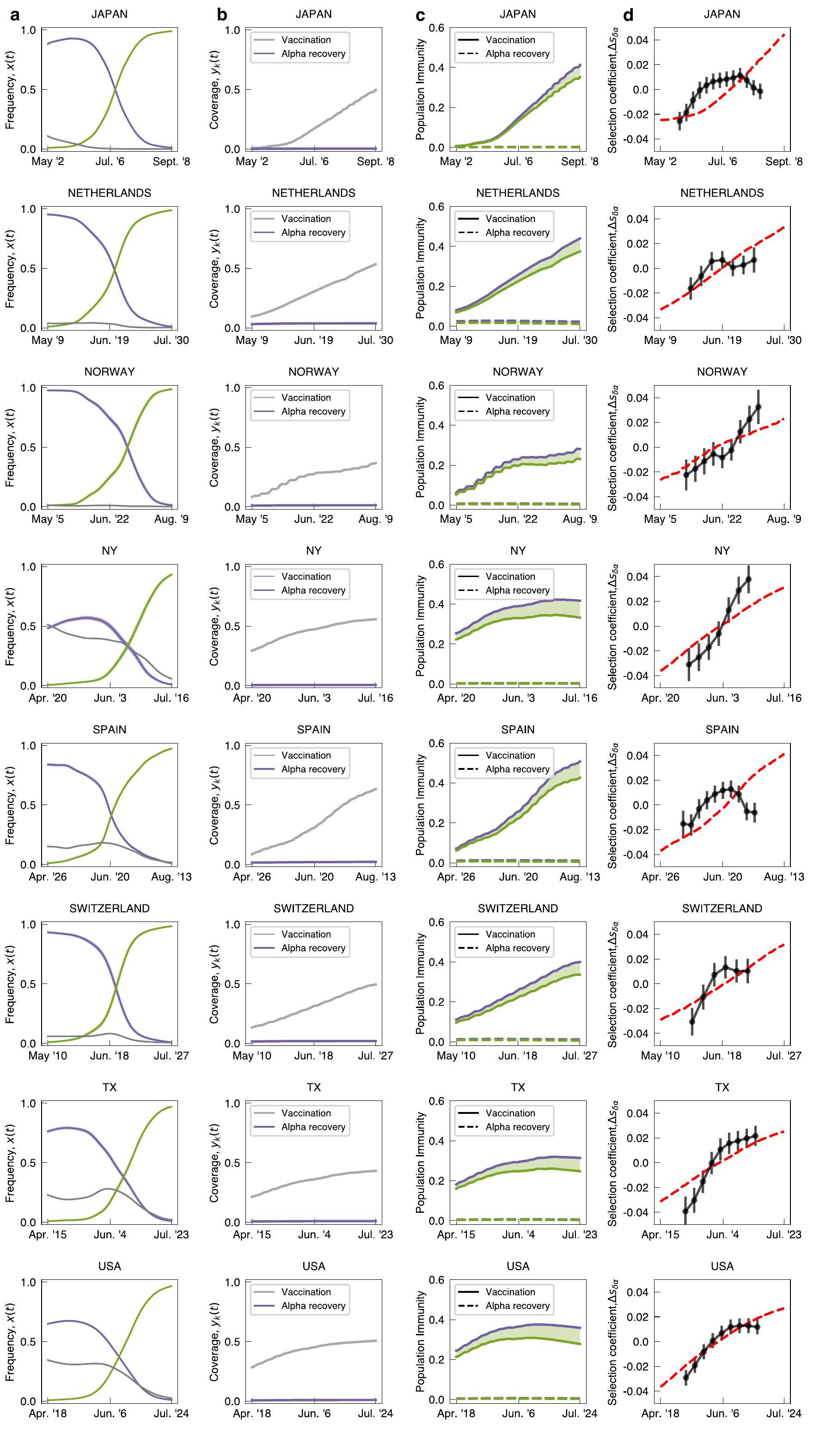}
\end{center}
\noindent 
{\footnotesize {\bf Fig. S1}: (continued) 
 }
\end{figure*}

\newpage
\begin{figure*}[t!]
\begin{center}
\includegraphics[width= 0.8 \linewidth]{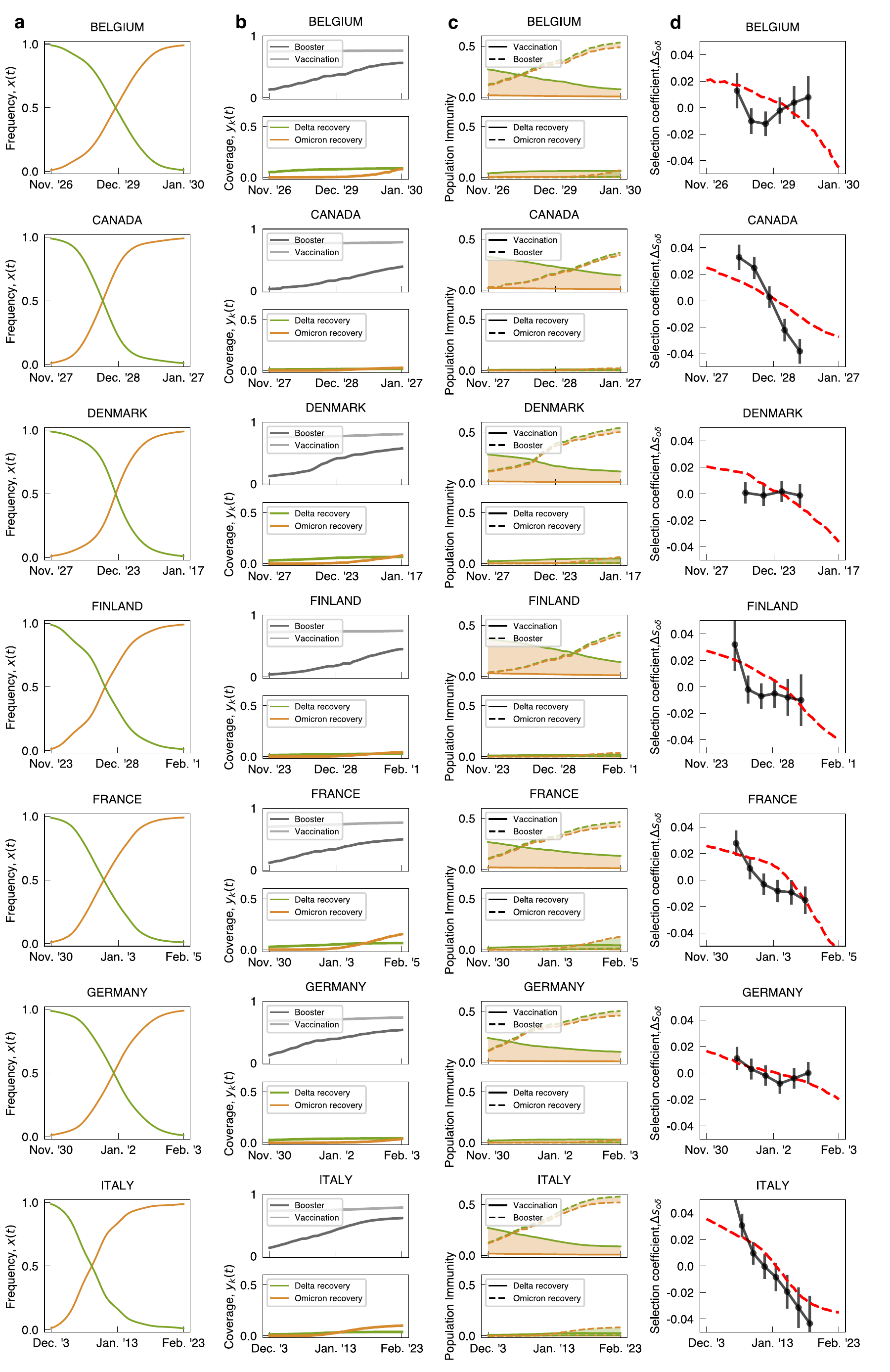}
\end{center}
\begin{flushright} 
\vspace*{-0.5cm}
{\footnotesize (continued on next page)}
\vspace*{-0.5cm}
\end{flushright}
\noindent 
\vspace*{-0.5cm}
\caption*{\footnotesize {\bf Fig. S2}:
{\bf Empirical and model-based trajectories of the $\delta - o$ shift.}  Evolutionary, epidemiological, and cross-immune trajectories are shown for all regions of this study. 
 {\bf (a)}~Observed frequency trajectories of relevant clades, $x_i (t)$; rms sampling error is indicated by shading.
 {\bf (b)}~Cumulative coverage of primary vaccination, $y_\vac (t)$ (light gray), and of booster vaccination, $y_\bst (t)$ (dark gray); cumulative population fraction of $\delta$ infections, $y_\delta (t)$ (green), and of $o$ infections, $y_o (t)$ (orange). 
  {\bf (c)}~Population immunity functions, $C_i^k (t)$ (as in Fig.~1c). 
 {\bf (d)}~Empirical selection change, $\Delta \hat s (t)$ (dots, with rms statistical errors indicated by bars), together with ML model prediction, $\Delta s (t)$ (dashed line). Criteria for inclusion of regions are given in Methods. 
}
\label{fig:S3}
\end{figure*}

\newpage
\begin{figure*}[t!]
\begin{center}
\vspace*{-1.6cm} 
\includegraphics[width= 0.8 \linewidth]{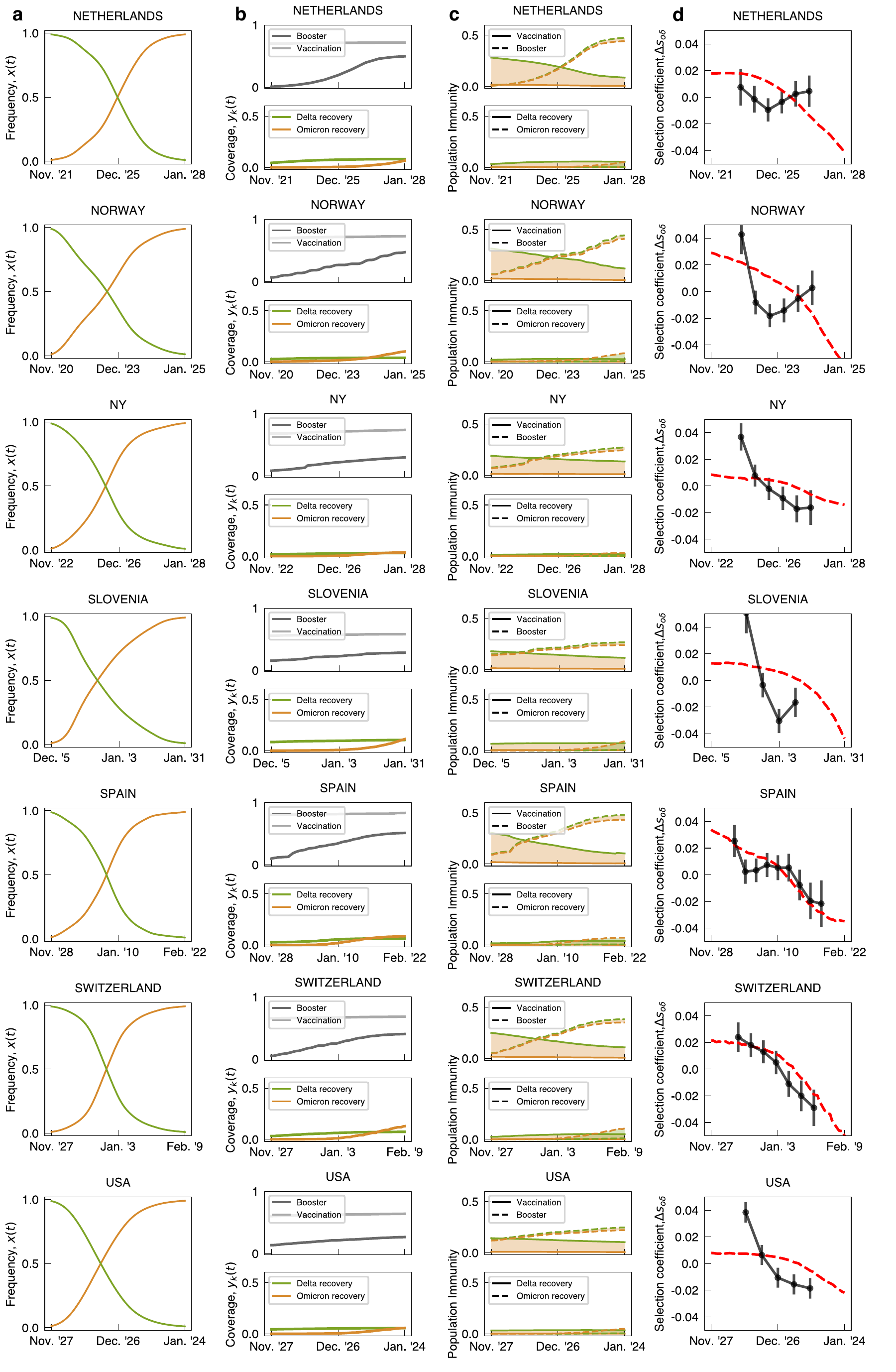}
\end{center}
\noindent 
{\footnotesize {\bf Fig. S2}: (continued) 
 }
\end{figure*}

\newpage
\begin{figure*}[t!]
\centering
\includegraphics[width= \linewidth]{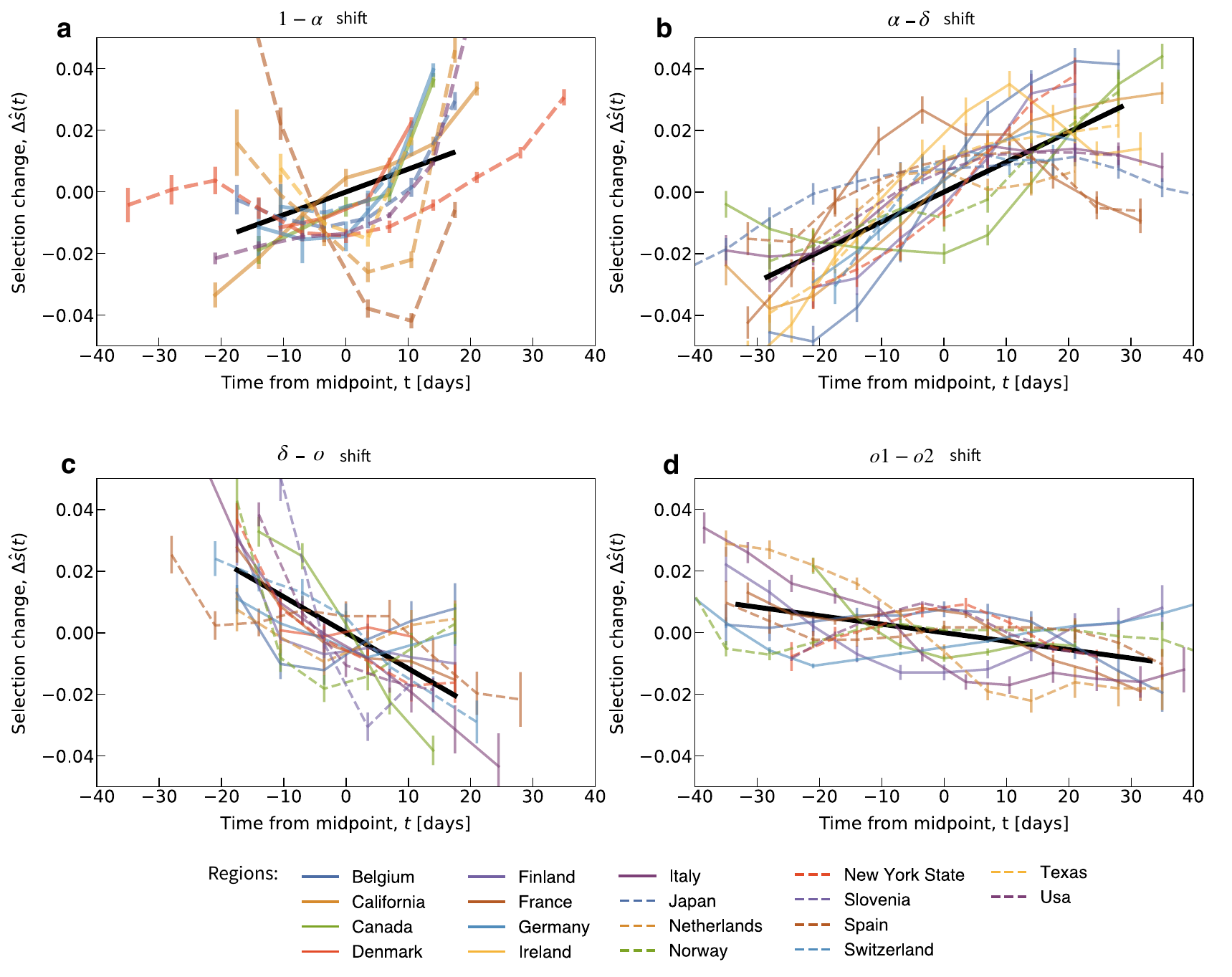}
\noindent 
\caption*{\footnotesize {\bf Fig. S3}:
{\bf Selection tracking in multiple regions and clade shifts.}  
Empirical selection change between invading and ancestral clade, $\Delta \hat s (t) = \hat s (t) - \langle s \rangle$, for all complete clade shifts and all regions of this study (brackets denote time averages for each trajectory). Selection trajectories are derived from  regional frequency trajectories and plotted against time counted from the midpoint (colored lines); rms statistical error is indicated by shading.
Summary statistics: cross-region linear regression, $s_{\rm lin} (t)$ (black solid line, length gives r.m.s.~time span of trajectories). 
{\bf (a)} $\anc - \alpha$ shift: small, statistically insignificant time dependence, ${\rm Var} (s_{\rm lin}) = 3.6 \times 10^{-4}$, $P > 0.01$;
{\bf (b)} $\alpha - \delta$ shift: substantial, statistically significant time dependence, ${\rm Var} (s_{\rm lin}) = 2. \times 10^{-3}$, $P < 10^{-16}$;
{\bf (c)} $\delta - o$ shift: substantial, statistically significant time dependence, ${\rm Var} (s_{\rm lin}) = 1.4 \times 10^{-3}$, $P < 10^{-5}$;
{\bf (a)} $o1 - o2$ shift: small, but statistically significant time dependence, ${\rm Var} (s_{\rm lin}) = 2.9 \times 10^{-4}$, $P < 10^{-4}$. All $P$ values are computed  using a two-sided Wald test.
The statistical grading of shifts is described and criteria for inclusion of regions are given in Methods. 
 }
\label{fig:XT}
\end{figure*}

\newpage 
\begin{figure*}[t!]
\centering
\includegraphics[width= 1.0 \linewidth]{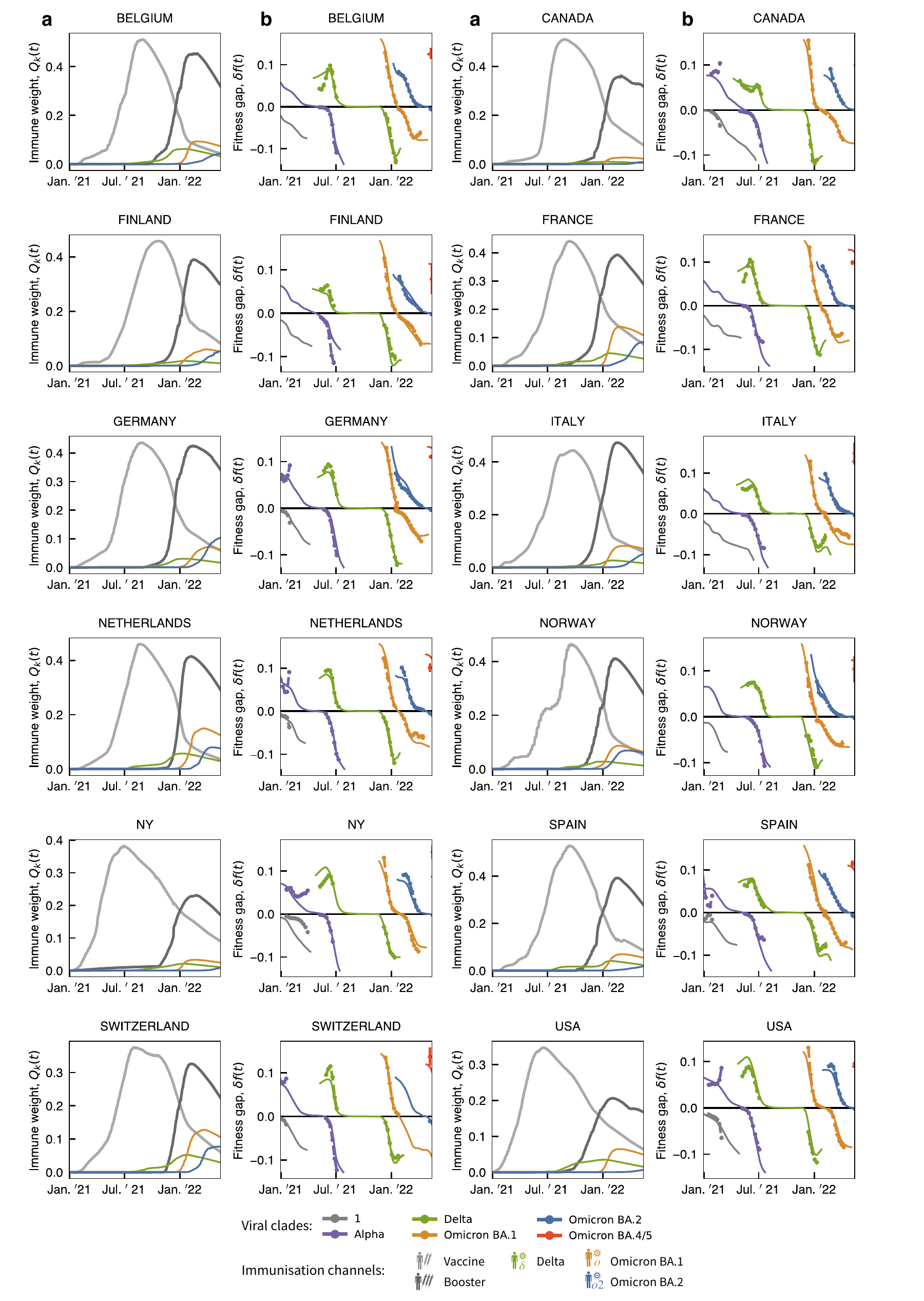}
\noindent 
\caption*{\footnotesize {\bf Fig. S4: Regional long-term trajectories of immune weight and fitness.} 
{\bf (a)} Time-dependent weight factors of different immune classes, $Q_k (t)$. 
{\bf (b)} Time-dependent fitness gap, $\delta f_i (t)$. 
Criteria for inclusion of regions are given in Methods; see Fig.~4 for averaged trajectories. 
}
\label{fig:LT}
\end{figure*}

\end{document}